\documentclass[prd,%
preprint,%
tightenlines,superscriptaddress,showpacs,%
nofootinbib,eqsecnum,amsfonts,amsmath]{revtex4}
\usepackage{bm}
\def\be{\begin{equation}}
\def\ee{\end{equation}}
\def\bea{\begin{eqnarray}}
\def\eea{\end{eqnarray}}

\begin{document}
\title{Dirac Quantization of Parametrized Field Theory}


\author{Madhavan Varadarajan}\email{madhavan@rri.res.in} 
\affiliation{Raman Research
Institute, Bangalore 560 080, India}

\begin{abstract}

Parametrized field theory (PFT) is free field theory on flat spacetime in a 
diffeomorphism invariant disguise. It describes field evolution on {\em 
arbitrary} (and  in general, curved) foliations of the flat spacetime instead 
of only the usual flat foliations, by treating the `embedding variables' which 
describe the foliation as dynamical variables to be varied in the action in 
addition to the scalar field. A formal Dirac quantization turns the constraints
of PFT into functional Schrodinger equations which describe evolution of 
quantum states from an arbitrary Cauchy slice to an infinitesimally nearby one.
This formal Schrodinger picture- based quantization is unitarily equivalent to 
the standard Heisenberg picture based Fock quantization of the free scalar 
field if scalar field  evolution along arbitrary foliations is unitarily 
implemented on the Fock space. Torre and Varadarajan (TV) showed that for 
generic foliations emanating from a flat initial slice in spacetimes of 
dimension greater than 2,  evolution is not unitarily implemented, thus 
implying an obstruction to Dirac quantization.

We construct a Dirac quantization of PFT, unitarily equivalent to the standard 
Fock quantization, using techniques from Loop Quantum Gravity (LQG) which are 
powerful enough to super-cede the no- go implications of the TV results. 
The key features of our quantization include an LQG type representation for the
 embedding variables, embedding dependent Fock spaces for the scalar field, an 
anomaly free representation of (a generalization of) the finite 
transformations generated by the constraints and group averaging techniques. 
The difference between the 1+1 dimensional case and  the case  of higher 
spacetime dimensions is that for the latter, only finite gauge transformations
are defined in quantum theory not the infinitesimal ones.

\end{abstract}
\maketitle


\section{Introduction}
 The usual description of quantum field theory on flat spacetime employs inertial coordinates. Such a description
leads, naturally, to a study of quantum dynamics from one instant of inertial time to another. Each such instant
corresponds to a flat spacelike Cauchy slice of the spacetime and the quantum dynamics evolves quantum fields
between  two such inertial slices. In contrast, in this work, we study aspects of quantum evolution of a free scalar
field between two  {\em arbitrary} (and in general, curved) Cauchy slices of the flat spacetime.

An elegant way to describe classical evolution of a free scalar field along arbitrary foliations of flat spacetime 
is through a formulation known as Parameterized field theory.
Parametrized field theory (PFT) is free field theory on flat 
spacetime in a diffeomorphism invariant disguise \cite{kareliyer}. 
Due to its general covariance and solvability, Kucha{\v r} pioneered the
use of PFT as a toy model to study various issues which arise in canonical quantum gravity 
(see \cite{kareliyer,karel1+1c,karel1+1q} and references therein). 
Our interest in PFT is, likewise, in its use as a toy model for quantum gravity.

PFT describes field evolution along  arbitrary 
foliations of the flat spacetime 
by treating the `embedding variables' which describe 
the foliation as dynamical variables to be varied in the action 
in addition to the scalar field. 
Let $X^{A}:= (T,X^1,..,X^n)$ denote inertial coordinates on an $n+1$ dimensional flat spacetime.
In PFT, $X^{A}$ are parametrized by a new set of arbitrary coordinates 
$x^{\alpha}=(t,x^1,..,x^n)$ such that  for fixed  $t$,  the `embedding variables' $X^{A}(t,x^1,..x^n)$ 
define a spacelike Cauchy slice of the flat  spacetime.
General covariance of PFT ensues from the arbitrary choice of $x^{\alpha}$ 
and implies that in its canonical description, evolution from one slice of 
an arbitrary foliation to another is generated by constraints. 

The constraints of PFT in its Hamiltonian formulation take the form
\be
{C}_{A}(x) := P_{A}(x) 
+ h_{A}[\phi,\pi,X^{B}](x) = 0 \;\;,
\label{ca}
\ee
where $P_{A}$ and $\pi$ are the momenta conjugate to $X^{A}$ 
and $\phi$, and $h_{A}$ is related to the stress-energy of 
the scalar field. If we formally define  ${\hat X^{A}}$ to act by multiplication
and ${\hat P}_{A}$ by functional differentiation, we may attempt to construct a  
Dirac quantization 
in which the  formal operator 
version of $C_{A}$ acting on a physical state 
$\arrowvert \Psi \rangle$ of the theory is given by
\be
\left(\frac{1}{i}\frac{\delta}{\delta X^{A}} 
+ \hat{h}_{A} \right) \arrowvert \Psi \rangle = 0 \;\;.
\label{fsch}
\ee
Eq.(\ref{fsch}) takes the form of a functional Schr\"odinger equation 
which represents infinitesimal evolution of the quantum state 
$\arrowvert \Psi \rangle$ from one Cauchy slice to another.
The question of interest in this paper is whether a Dirac quantization of PFT yields
a theory which is unitarily equivalent to the standard Fock quantization of the 
free scalar field on flat spacetime.

Note that the standard Fock representation is in the Heisenberg picture because 
the field operators are time dependent 
and the states, time independent. In contrast, the representation used in (\ref{fsch}) 
is one in which the operators ${\hat \phi}, {\hat \pi}$ commute with ${\hat X^{A}}, 
{\hat P}_{A}$ (in accordance with their classical Poisson brackets) and hence are 
embedding -independent, whereas the states are embedding dependent.  
Thus, Dirac quantization leads to a Schrodinger picture based representation. The Schrodinger and 
Heisenberg pictures are unitarily related iff the evolution of the scalar field operators from
some fixed initial  Cauchy slice (this slice being the analog of the initial instant of time used to define the 
Schrodinger picture in quantum mechanics), to an arbitrary final Cauchy slice 
is unitarily implemented in the 
Heisenberg picture. If such evolution is unitary, the Schrodinger picture can be defined as the 
`back- evolved' image of the Heisenberg picture. 

Torre and Varadarajan (TV) \cite{tv1}
showed that in 1+1 dimensions  operator evolution is unitary, 
the functional Schrodinger picture exists as the unitary image of the standard Heisenberg 
picture based Fock quantization,
 and the functional Schrodinger equation (\ref{fsch}) is rigorously 
defined. However, for  spacetimes of  dimension greater than 2 and for generic choices of the final Cauchy slice, 
TV found  that operator evolution
from a flat initial slice is not  unitarily implementable \cite{tv2} .
Thus the TV results seem to indicate an obstruction to a physically correct 
Dirac quantization of higher dimensional PFT. If a Dirac quantization is not viable even in this 
simplest of models, it is a matter of concern for any Dirac quantization based approach to quantum gravity.

Recent results of Cho and Varadarajan \cite{chome} reinforce these apprehensions. They analyse the 
case of 
an axisymmetric scalar field evolving along arbitrary axisymmetric
foliations of a flat  2+1 dimensional spacetime. This system is equivalent, via Kucha{\v r}'s canonical 
transformation \cite{karelcyl}, to the midisuperspace of cylindrically symmetric (1 polarization) gravitational
fields. Reference \cite{chome} obtains a non- unitary result for generic axisymmetric scalar field evolution 
by showing that  the action of generic
radial diffeomorphisms on the scalar field operators is not unitarily implementable (see \cite{chome} for details).

One of the most active  approaches to a Dirac quantization of gravity is Loop Quantum Gravity (LQG). 
We are particularly interested in the implication of the
results of \cite{tv2,chome} for LQG. Indeed, since a unitary representation of spatial diffeomorphisms lies at the heart of 
LQG, the results of \cite{chome} seem to be of significant concern. 
In this work, we show that these apprehensions are unjustified by constructing a Dirac quantization of PFT (in 
any spacetime dimension) which is unitarily equivalent to the standard Fock quantization. Our construction
makes vital use of techniques developed in LQG (see \cite{alm2t} and references therein). 
Indeed, in the absence of these techniques, this work would not have
been possible. 

How do we avoid the negative implications of references \cite{tv2,chome}? The key point is that LQG methods demand
a Hilbert space representation for {\em all} the phase space variables i.e. both the scalar field and the 
embedding variables. The arguments leading to Eq. (\ref{fsch}) were based on a heuristic treatment of the 
embedding variables; ${\hat X}^{A}, {\hat P}_{A}$ were not defined as operators on a {\em Hilbert} space.
As we shall see, once the correct  Hilbert space representation is defined for the embedding variables,
it is possible to construct a viable Dirac quantization. Specifically, we define an LQG type of representation on a non- seperable
Hilbert space for the embedding sector. This enables us to specify  embedding- dependent Fock space representations
for the matter sector. The resultant non- seperable Hilbert space for the embedding and the matter field operators is called
the kinematic Hilbert space in LQG terminology. It
provides an anomaly free, unitary representation
for a certain generalization of 
the {\em finite} (as opposed to infinitesmal) transformations generated by the constraints. Using Group Averaging 
techniques from LQG \cite{alm2t,dongrpavg} one can construct physical states from kinematic ones. The physical Hilbert 
space,
${\cal H}_{phys}$,
contains  those
states which are invariant under the unitary action of the `exponentiated' constraints and comes with a natural
inner product as a result of Group Averaging. It can then easily be seen that the resulting representation of 
Dirac observables is such that the theory is unitarily equivalent to the standard Fock quantization.
Although the construction is valid for any dimension, it turns out that in higher dimensions only
the finite transformations generated by the
constraints  (more precisely, their generalization, referred to above,) can be defined
in the quantum theory. This is in contrast to the 1+1
case where, in a precise sense, infinitesmal transformations can also be defined.

We proceed as follows.  In section 2 we provide a brief review of PFT. In section 3, we specify the representation 
for the embedding variables. In section 4, we define the representation for the matter variables. This
completes the specification of the kinematic Hilbert space.  In section 5 we show that there is  a problem with 
our choice of operators depending on the embedding momenta 
\footnote{This problem can be traced to the fact that  
the requirement that each embedding must define a spacelike slice is contradiction with a vector space structure on 
the space of embeddings. It is the analog of the problem of preserving the non- degeneracy of the spatial metric
in quantum geometrodynamics \cite{ishamnondeg}.\label{f1}} 
and arrive at an improved choice of operators. These operators are the 
quantum  correspondents of the finite canonical transformations generated by the constraints. We show that these operators
have a unitary action on the kinematic Hilbert space. Next, in section 6, we complete the Dirac quantization by constructing the
physical Hilbert space through Group Averaging. We display the action of Dirac observables on ${\cal H}_{phys}$ and demonstrate 
unitary equivalence with the standard Fock representation. We also show how
to obtain the functional Schrodinger equation in 1+1 dimensions where a suitable infinitesmal version of the unitary action 
of section 5 can be defined. Section 7 is devoted to a discussion of our results. In particular we discuss 
the existence of inequivalent (Dirac) quantizations  in higher dimensions (this is the true implication of the TV results!)
and the presence of the  Virasoro anomaly in 1+1 dimensions even though the algebra of the unitary transformations of section 5
is anomaly free (this discussion is essentially a reminder of the  comments of Kucha{\v r} in \cite{karel1+1q} and of
 Kucha{\v r} and Torre in \cite{charliekarelstring}).

In what follows, we use units in which  $\hbar= c =1$.

\section{Review of Parametrized Field Theory.}
In this section we provide a brief review of PFT. The reader may consult \cite{kareliyer} for details.

\subsection{The Action for PFT.}
PFT is free scalar field theory on a fixed $n+1$ dimensional flat spacetime
written in a diffeomorphism invariant manner as follows. The action for a free scalar field $\phi$ on
a fixed flat spacetime in terms of global inertial coordinates $X^A,\;A=0,..,n$ is 
\begin{equation}
S_0[\phi] = -\frac{1}{2} \int d^{n+1}X \eta^{AB}\partial_A\phi \partial_B\phi ,
\label{s0}
\end{equation}
where  the Minkowski metric 
in inertial coordinates, $\eta^{AB}$,  is diagonal with entries $(-1,1,1...1)$. If instead, we use coordinates 
$x^{\alpha}\;, \alpha= 0,..,n$ (so that $X^A$ are `parameterized' by $x^{\alpha}$,  $X^A= X^A(x)$), we have
\begin{equation}
S_0[\phi] = -\frac{1}{2} \int d^{n+1}x \sqrt{\eta} \eta^{\alpha \beta}\partial_{\alpha}\phi \partial_{\beta}\phi ,
\label{s0x}
\end{equation}
where $\eta_{\alpha \beta}= \eta_{AB} \partial_{\alpha}X^A\partial_{\beta}X^B$ and $\eta$ denotes the 
determinant of $\eta_{\alpha \beta}$. The action for PFT is obtained 
by considering the right hand side of (\ref{s0x}) as a functional, not only of $\phi$, but also of 
$X^A(x)$ i.e. $X^A(x)$ are considered as  $n+1$ new scalar fields to be varied in the action ($\eta_{\alpha \beta}$
is a function of $X^A(x)$). Thus 
\begin{equation}
S_{PFT} [\phi, X^A]= -\frac{1}{2} \int d^{n+1}x \sqrt{\eta(X)} \eta^{\alpha \beta}(X)\partial_{\alpha}\phi \partial_{\beta}\phi .
\label{spft}
\end{equation}
Note that $S_{PFT}$ is a diffeomorphism invariant functional of the scalar fields $\phi (x), X^A (x)$.
Variation of $\phi$ yields the equation of motion $\partial_{\alpha}(\sqrt{\eta}\eta^{\alpha \beta}\partial_{\beta}\phi)= 0$,
which is just the flat spacetime equation $\eta^{AB}\partial_A\partial_B \phi=0$ written in the coordinates $x^{\alpha}$.
On varying $X^A$, one obtains equations which are satisfied if  $\eta^{AB}\partial_A\partial_B \phi=0$.
This implies that $X^A(x)$ are $n+1$  undetermined functions (subject to the condition that determinant of $\partial_{\alpha}X^A$
is non- vanishing). This $n+1$- function- worth of gauge is a reflection of the $n+1$ dimensional diffeomorphism invariance
of $S_{PFT}$. Clearly the dynamical content of $S_{PFT}$ is the same as that of $S_0$; it is only that 
the diffeomorphism invariance of 
$S_{PFT}$ naturally allows a description of the standard free field dynamics dictated by $S_0$ on {\em arbitrary}
foliations of the fixed flat spacetime.

\subsection{Hamiltonian Formulation of PFT.}
In the previous subsection, $X^{A}(x)$ had a dual interpretation - one as  dynamical variables to be varied in the action, and 
the other as inertial coordinates on a flat spacetime. In what follows we shall freely go between these two interpretations.

We set $x^0=t$ and $\{x^{\alpha}\} = \{t, x^a,\;a =1..n\}$. We restrict attention to $X^A(x)$ such that for any fixed 
$t$, $X^A(t,x^a)$ describe an embedded spacelike hypersurface in the $n+1$ dimensional flat spacetime (it is for this
reason that $X^A(x)$ are called embedding variables in the literature). This means that, for fixed $t$, the functions $X^A(x)$
must be such that the symmmetric form $q_{ab}$ defined by 
\be
q_{ab}(x):=\eta_{AB}\frac{\partial X^A(x)}{\partial x^a}\frac{\partial X^B(x)}{\partial x^b}
\label{nondeg}
\ee
is an $n$- dimensional Riemannian metric. This follows from the fact that $q_{ab}(x)$ is the induced metric 
on the hypersurface in the flat spacetime defined by $X^A(x)$ at fixed $t$.

An $n+1$ decomposition of $S_{PFT}$ with respect to the time `$t$', leads to its Hamiltonian form:
\begin{equation}  
S_{PFT} [\phi,X^A;\pi,P_A;N^A]
=\int dt \int d^{n}x (P_A{\dot X}^A + \pi {\dot \phi}- N^A C_A).
\label{spftham}
\end{equation}
 Here $\pi$ is the momentum conjugate to the scalar field $\phi$, $P_A$ are the 
momenta conjugate to the embedding variables $X^A$, $N^A$ are Lagrange multipliers for the first class
constraints $C_A$, with $\{C_A,C_B\}=0$. The constraints are functions on phase space which are linear in the embedding momenta,
$C_A:= P_A - h_A (X,\phi,\pi)$, where $h_A$  are related to certain components of the stress energy tensor of the 
scalar field on the $t=$constant spatial hypersurface $X^A(t,x^a)$. It turns out that the motions on phase space 
generated by the `smeared'  constraints, $\int d^nx (N^AC_A)$ correspond to  
scalar field evolution along arbitrary foliations of the flat spacetime, each choice of foliation being in correspondence with 
a choice of multipliers $N^A$. Since the constraints are first class they also generate gauge transformations,
and as in General Relativity, the notions of gauge and evolution are intertwined. 

For future reference, we note  that if we choose to smear the constraints with functions of the embedding variables, 
the Poisson bracket algebra of the smeared constraints is isomorphic to the Lie algebra of diffeomorphisms on the 
spacetime manifold. Denote the spacetime manifold by $M$. Unless otherwise specified, we shall assume that $M$ is 
diffeomorphic to $R^{n+1}$. Consequently,
we shall also assume, unless otherwise specified, that the Cauchy slices  are diffeomorphic to $R^n$ so that 
$x^a$ are global coordinates on $R^n$. We refrain from specifying boundary conditions/asymptotic conditions
 of various phase space variables
and smearing functions; we anticipate that this can easily be done and will not alter any of our results.

Let $\xi_1^A,\xi_2^A $  be two vector fields on $M$. We denote their dependence on the inertial 
coordinates $X^A$ by $\xi_I^A = \xi_I^A (X)\; I=1,2$. These vector fields define the two sets of functions of
the embedding variables, $\xi_I^A (X(x)),\; I=1,2$, in an obvious manner. We smear the constraints with these functions
to get $C(\xi_I)= \int d^nx (\xi_I^AC_A),\; I=1,2$. The smeared constraints have the algebra
\be
\{C(\xi_1), C(\xi_2)\} =  C({\cal L}_{\xi_2} \xi_1)
\label{cxipb}
\ee
where ${\cal L}_{\xi_2} \xi_1^A = 
\xi_2^B \frac{\partial\xi_1^A}{\partial X^B}-\xi_1^B \frac{\partial\xi_2^A}{\partial X^B}$.
Clearly, equation (\ref{cxipb}) displays an isomorphism with the Lie bracket between vector fields on $M$. The latter define the 
Lie algebra of diffeomorphisms on $M$ by virtue of the fact that the diffeormorphisms (connected to identity) on $M$ 
are generated by vector fields on $M$. 

Given a vector field $\xi^A$, we define, as before, the smeared constraint $C(\xi)$ by 
\be
C(\xi):= \int d^n x \xi^B(X(x)) C_B(x). 
\label{defcxi}
\ee 
Then it is easy to see that
\be
\{X^A(x), C (\xi )\}= \xi^A (X(x)) = \xi^B \frac{\partial X^A}{\partial X^B} =:{\cal L}_{\xi}X^A .
\label{cxiX}
\ee
In terms of structures on $M$, the above equation has the following interpretation.
$\xi$ generates a 1 parameter family of diffeomorphisms $d_{\xi}(t):M\rightarrow M$. Any diffeomorphism on $M$ 
maps the hypersurface  $X^A(x)$ to another hypersurface in $M$. The new embedding defined by the infinitesmal action
of $d_{\xi}(t)$ on $X^A(x)$ corresponds to that defined by equation (\ref{cxiX}) above.

It is also easy to see that 
\be
\{\phi(x) , C (\xi )\}= \{ \phi(x), \int d^nx \xi^A(X(x))h_A(x)\}, 
\label{cxiphi}
\ee
\be
\{\pi(x) , C (\xi )\}= \{ \pi(x), \int d^nx \xi^A(X(x))h_A(x)\} .
\label{cxipi}
\ee
It is straightforward to verify that the right hand side of the above equations has the following interpretation in terms
of free scalar field evolution on $M$. Let $(\phi (x), \pi (x))$ be initial data on the Cauchy  slice defined by 
$X^A (x)$. Consider the evolution of this data, via the free scalar field equations on $M$, to the infinitesmally nearby slice defined
by equation (\ref{cxiX}). The evolved data correspond to that defined by the right hand sides of (\ref{cxiphi}) and (\ref{cxipi}) above.

\subsection{The standard Fock representation via gauge fixing}
A gauge fixing of the constraints which maps the classical theory directly onto the standard description 
of a free scalar field in inertial coordinates is, not surprisingly,  $X^1=x^1,....,X^n=x^n, X^0=t$.
\footnote{Technically, this is not a gauge fixing but a 1 parameter family of gauge fixings (one for 
each value of $t$) known as `deparameterization'.\label{f2}} 
It is straightforward to implement this gauge fixing,
eliminate the embedding momenta and obtain a reduced action identical to  (the Hamiltonian form of)
 $S_0$ (\ref{s0}).  As is well known, this form of the theory admits the standard Fock quantization
wherein  quantum states $|\Psi>$ are time independent and the field operators ${\hat \phi}(x,t)$ are 
time dependent (note that,
here, ($x,t$) are chosen to be inertial coordinates by virtue of the gauge fixing conditions).

It is also possible to obtain the standard Fock quantization by first subjecting the canonical coordinates
$X^A, P_A, \phi, \pi$ to a Hamilton- Jacobi type of canonical transformation and then constructing a Dirac quantization 
\cite{kareliyer,karel1+1q}. In this work we are interested in a direct Dirac quantization of PFT without 
any canonical transformations to simplify the constraints. This is because we are interested in PFT as
a toy model for gravity and no such simplifying canonical transformation is known for gravity.

\section{The representation of the embedding variables}

We construct the representation for the embedding variables through the following steps.
First we specify a complete set of functions  on the embedding phase space which is closed 
under Poisson brackets and under complex conjugation. By complete, it is meant that any 
other function on this phase space can be written as the (limit) of sums and products of these functions.

Next, we define a representation of these functions as quantum operators so that the Poisson brackets go to 
quantum commutators. Finally, we specify an inner product on the representation space such that the relations implied 
by complex conjugation are enforced as adjointness relations on the operators.

As is conventional in Hamiltonian theory, we shall drop the explicit dependence of various variables on the time coordinate $t$.
Thus, the notation $f(x)$ signifies the functional dependence of the function $f$ on the coordinates $x^a$.

\subsection{The classical Poisson bracket algebra}
 
Let $G^A (x),\; A=0,..,N$ be  smooth functions of $x^a, \;a=1,..,n$. Define the functional, $H_G$, of the embedding momenta 
as follows:
\be
H_{G} [P_A] = \exp i \int d^nx G^B P_B .
\label{defhg}
\ee
We shall choose our complete set of functions as $(X^A (x), H_{G}[P_A])$.
It is easy to check that the only non- trivial Poisson brackets are
\be
\{ X^A (x), H_{G}\} = i G^A(x) H_G
\label{pbxhg}
\ee
The set of functions is also closed under complex conjugation since
\be 
(X^A(x))^* = X^A(x) \;\; (H_G)^* = H_{-G}
\label{rc}
\ee

\subsection{The Representation}
We specify the representation of our set of functions as quantum operators by defining their action on
basis states of our representation. Each state in the basis is labelled by
a set of smooth functions of $x^a, \; a=1,..,n$. Given the set of smooth functions, $F^A (x)\; A=0,..,N$, we denote the 
 corresponding basis state by $\arrowvert F \rangle$. We define the action of ${\hat X}^A, {\hat H}_G$ through
\be
{\hat X}^A (x)\arrowvert F \rangle = F^A (x)\arrowvert F \rangle,
\label{xhat}
\ee
\be
{\hat H}_G \arrowvert F \rangle = \arrowvert F-G \rangle .
\label{phat}
\ee
It is easy to verify that the above equations provide a representation of the classical Poisson bracket relations.

\subsection{Inner Product}

Given the smooth functions $F_1^A (x), F_2^A(x)$, we define the {\em Kronecker} delta function $\delta_{F_1,F_2}$ by 
\bea
\delta_{F_1,F_2} & =& 0 \;{\rm if\; there\; exists}\; A,x^a \;{\rm such\; that}\;  F_1^A (x)\neq  F_2^A(x) \nonumber \\ 
&= &  1 \;{\rm if}\; F_1^A (x) =  F_2^A(x) \;{\rm for \; all}\; A,x^a .
\label{deltaf1f2}
\eea
Then the  inner product between two states in our basis labelled by the smooth functions $F_1^A (x), F_2^A(x)$ is defined to be
\be
\langle F_1  \arrowvert F_2 \rangle =\delta_{F_1,F_2} .
\label{embedip}
\ee
Note that this inner product implies that our basis states are orthonormal. Since the basis states are uncountable,
we have a non- seperable representation space. Also note that, with this inner product, ${\hat H}_G$ does not have 
the appropriate continuity in $G^A(x)$ to define ${\hat P}_A(x)$; thus although ${\hat H}_G$ is a well defined operator,
this representation does not allow the existence of ${\hat P}_A(x)$.

It suffices to check the implementation of the `reality conditions' (\ref{rc}) in the context of pairs of basis states i.e.
it is straightforward to verify that 
\bea 
(\langle F_1  \arrowvert {\hat X}^A(x) F_2 \rangle )^* &=& \langle F_2  \arrowvert {\hat X}^A(x) F_1 \rangle ,
\label{rcx}
\\
(\langle F_1  \arrowvert {\hat H}_G F_2 \rangle )^* &=& \langle F_2  \arrowvert {\hat H}_{-G} F_1 \rangle .
\label{rchg}
\eea
For our purposes in subsequent sections we could stop here. However, for completeness, we specify the obvious steps
for the construction of the Hilbert space and complete verification of the reality conditions thereon.

Define the set ${\cal D}_X$ to be the set of finite linear combinations of the basis states defined above. Thus
$\arrowvert \psi \rangle \in {\cal D}_X$ iff
\be
\arrowvert \psi \rangle = \sum_{I=1}^N a_I \arrowvert F_I \rangle ,
\label{defdense}
\ee
where $F_I^A (x),\; I=1,..,N$ are smooth functions such that $F_I^A(x) = F_J^A(x)$ only when $I=J$, 
$N$ is finite and $a_I, \;I=1,..,N$ are complex numbers.
The inner product (\ref{embedip}) can be extended to ${\cal D}_X$ by appropriate linearity.
${\cal D}_X$ serves as a dense set for the embedding sector Hilbert space ${\cal H}_X$ i.e. the Cauchy completion
of  ${\cal D}_X$ in the inner product (\ref{embedip}) yields  ${\cal H}_X$. Thus 
$\arrowvert \psi \rangle \in {\cal H}_X$ iff
\be
\arrowvert \psi \rangle = \sum_{I=1}^\infty a_I \arrowvert F_I \rangle ,
\label{defhilbertx}
\ee
where $F_I^A (x)$ are smooth functions such that $F_I^A(x) = F_J^A(x)$ only when $I=J$ and 
$\sum_{I=1}^{\infty} |a_I|^2 <\infty$.
Note that ${\cal D}_X$ serves as a dense domain for the unbounded operator ${\hat X}^A(x)$.

\section{The Representation for the scalar field variables}

\subsection{Preliminary Remarks}
The only non- trivial Poisson bracket involving $\phi(x)$ and/or $\pi (x)$ is 
$\{\phi (x), \pi (y)\} = \delta (x,y)$. In particular we have the trivial Poisson brackets
$\{\phi (x), P_A (y)\}= 0$ and $\{\pi (x), P_A (y)\}= 0$ which are equivalent to the equations
$\frac{\delta \phi (x)}{\delta X^A (y)}=0$ and $\frac{\delta \pi (x)}{\delta X^A (y)}=0$.
As a result of the latter equations, we seek, in quantum theory, a representation of
${\hat \phi} (x),{\hat \pi} (x)$ which is independent of the embeddings. This is like the 
Schrodinger picture in standard quantum mechanics,  with (the eigenvalues of)
${\hat X}^A (x)$ playing the role of time. Bearing this analogy in mind, we shall pattern our constructions
below on corresponding  structures in standard quantum mechanics.

In usual particle quantum mechanics, the Schrodinger picture is constructed by evolving the Heisenberg 
picture operators at time $t$ back to some fixed initial time $t_0$, usually chosen to be $t_0=0$. Here, 
in order to make contact with \cite{tv1,tv2} we choose the analog of $t_0=0$ to be the initial flat
embedding $F_0^A(x) = (0,x^1,..,x^n)$. Thus, we want ${\hat \phi} (x),{\hat \pi} (x)$  to act as if they were Heisenberg picture fields
at the initial embedding $F_0^A(x)$. This prompts the definitions
\be
{\hat \phi (x)}= 
\frac{1}{(2\pi)^{\frac{n}{2}}}\int \frac{d^nk}{\sqrt{2k}}
 ({\hat a}_S(k) e^{i{\vec k}\cdot{\vec x}} + {\hat a}^{\dagger}_S(k) e^{-i{\vec k}\cdot{\vec x}}),
\label{defphis}
\ee
\be
{\hat \pi (x)}= 
\frac{1}{(2\pi)^{\frac{n}{2}}}\int \frac{d^nk\sqrt{k}}{\sqrt{2}}
 (-i{\hat a}_S(k) e^{i{\vec k}\cdot{\vec x}} + i{\hat a}^{\dagger}_S(k) e^{-i{\vec k}\cdot{\vec x}}),
\label{defpis}
\ee
where we have denoted $(x^1,...,x^n)$, $(k^1,..,k^n)$ by ${\vec x}$, ${\vec k}$ 
and where $k:=\sqrt{{\vec k}\cdot{\vec k}}$ 
If evolution is unitary, the Schrodinger and Heisenberg picture representations exist on the same Hilbert space.
If this were the case, we could define ${\hat a}_S(k), {\hat a}^{\dagger}_S(k)$ in the usual
way as annihilation and creation operators. We could then generate the Fock space by the action of the 
creation operators on the standard Fock vacuum. Since for spacetime dimensions greater than 2 (i.e. for $n>1$)
generic evolution is not unitary \cite{tv2}, this is {\em not} the case and, instead, we define the 
action of  ${\hat a}_S(k), {\hat a}^{\dagger}_S(k)$ in analogy with the following structure in usual quantum mechanics.

\subsection{The Heisenberg and Schrodinger pictures in standard quantum mechanics.}
Consider a system with phase space coordinatized by (complex) canonical coordinates $(a, a^*)$ so that
$\{ a, a^* \}=-i$. Let its dynamics be
described by some time dependent quadratic Hamiltonian. Consequently,
classical evolution is a linear canonical 
transformation so that 
\be
a (t_2) = \alpha (t_2, t_1)  a(t_1) + \beta(t_2, t_1) a^*(t_1)
\label{at1t2cm}
\ee
 where $\alpha, \beta$ are complex functions (of the initial and 
final times $t_1$ and $t_2$) which are determined by the Hamiltonian of the system.
Denoting the pair $a, a^*$ by $\vec{a}$ and defining the matrix $C(t_2, t_1)$ by 
\be
C(t_2, t_1) =
\left(
\begin{array}{cc}
\alpha & \beta\\           
\beta^* & \alpha^*
\end{array}
\right),
\label{cacm}
\ee
we can write equation (\ref{at1t2cm}) as 
\be
\vec{a}(t_2) = C(t_2, t_1) \vec{a} (t_1).
\label{veccm}
\ee
For future purposes we note that  equation (\ref{veccm}) implies  the  identities:
\be 
C(t_3, t_1)=C(t_3, t_2)C(t_2, t_1), \;\;\;\;\; C(t,t)= {\bf 1},
\label{cmultqm}
\ee
where $\bf 1$ denotes the identity operator.
In the Heisenberg picture, $a(t),a^*(t)$ are represented by the operators
${\hat a}_H(t),{\hat a}_H^{\dagger}(t)$. The time independent state space is generated by the 
action of the creation operator, ${\hat a}_H^{\dagger}(0)$, on the vacuum $\arrowvert 0\rangle$, the 
vacuum being defined by ${\hat a}_H(0)\arrowvert 0\rangle$= 0. Clearly, ${\hat a}_H(t),{\hat a}_H^{\dagger}(t)$
are related to the $t=0$ operators through the relation 
\be
\vec{\hat a}_H(t) = C(t, 0)\vec{\hat a}_H(0),
\label{ctoqm}
\ee
where we have used the obvious notation that $\vec{\hat a}_H(t)$ denotes the pair ${\hat a}_H(t),{\hat a}_H^{\dagger}(t)$.

Let the unitary transformation corresponding to evolution from $t_1$ to $t_2$ be denoted by 
$U(t_2, t_1)$. Then the Schrodinger picture representative of ${\vec a}(t)$ is given by
\be
\vec{\hat a}_S = U^{\dagger}(t,0) \vec{\hat a}_H(t) U(t,0) = \vec{\hat a}_H(0)
\ee
and the Schrodinger picture `vacuum' is given by 
\be
\arrowvert 0,t\rangle_S = U^{\dagger}(t,0)\arrowvert 0\rangle .
\ee
Clearly, the Schrodinger vacuum is annihilated by  the operator
\be
{\hat b}(t) = U^{\dagger}(t,0) \vec{\hat a}_H(0) U(t,0)=U^{\dagger}(t,0) \vec{\hat a}_S U(t,0)
\label{bqm}
\ee
and the Schrodinger picture image of the Heisenberg state $({\hat a}_H^{\dagger}(0))^n  \arrowvert 0\rangle$ is
the state $({\hat b}^{\dagger}(t))^n  \arrowvert 0,t\rangle_S$. Note that the Schrodinger picture image of 
equation (\ref{ctoqm}) is 
\be
\vec{\hat a}_H(0)= \vec{\hat a}_S = C(t,0) \vec{\hat b}(t), 
\label{abqm}
\ee
which, in conjunction with equation (\ref{cmultqm}), 
implies that 
\be
{\hat b}(t)= C(0,t) \vec{\hat a}_S .
\label{defbqm}
\ee
Given the operator $\vec{\hat a}_S$, the above equation can be defined even in the absence of the unitary tranformation
$U(t,0)$ and hence can be used to define the representation in the field theory case.

\subsection{Some considerations in classical field theory.}

Before we define the representation for the matter sector of PFT, it is useful to review the classical (field theoretic)
structures which are in correspondence with the ones in section 4B.

Consider free scalar field evolution from the  slice defined by the embedding $F^A_1(x)$ to the slice defined by the 
embedding $F^A_2(x)$. Let the scalar field data on $F_I^A(x),\; I=1,2$ be denoted by 
$(\phi (x), \pi (x)) = (\phi_{F_I} (x), \pi_{F_I} (x))$. Instead of the scalar field data, it is 
more convenient to work with the modes $a (\vec{k}), a^* (\vec{k})$ where
\be
a(\vec{k}) = 
\frac{1}{(2\pi)^{\frac{n}{2}}}\int d^nx \sqrt{\frac{k}{2}}
 (\phi (x) + i\frac{\pi (x)}{k})e^{-i{\vec k}\cdot{\vec x}}.
\label{defak}
\ee
Thus $a (\vec{k}), a^* (\vec{k})$ are the classical correspondents of ${\hat a}_S(\vec{k}),{\hat a}^{\dagger}_S(\vec{k})$
defined in equations (\ref{defphis}),(\ref{defpis}).

When the data for the scalar fields is $(\phi (x), \pi (x)) = (\phi_{F_I} (x), \pi_{F_I} (x))$, we denote the corresponding 
evaluation of the modes $a (\vec{k}), a^* (\vec{k})$ by $a_{F_I} (\vec{k}), a_{F_I}^* (\vec{k})$. Explicitly,
\be
a_{F_I}(\vec{k}) = 
\frac{1}{(2\pi)^{\frac{n}{2}}}\int d^nx \sqrt{\frac{k}{2}}
 (\phi_{F_I} (x) + i\frac{\pi_{F_I} (x)}{k})e^{-i{\vec k}\cdot{\vec x}}.
\label{defakfi}
\ee
As in  (\ref{at1t2cm}), classical evolution is a linear  canonical tranformation so that 
\be
a_{F_2} ({\vec k}) = \int d^nl\alpha_{F_2,F_1} (\vec{k},\vec{l}) a_{F_1} ({\vec l}) + 
\int d^nl \beta_{F_2,F_1}(\vec{k},\vec{l}) a_{F_1}^* ({\vec l}).
\label{af1f2}
\ee
The coefficients $\alpha, \beta$ satisfy the Bogoliubov conditions \cite{bogbovcondtns} by virtue of the transformation being 
canonical. In analogy to section 4B, we denote the pair of functions
$a_{F_I} (\vec{k}), a_{F_I}^* (\vec{k})$ by ${\vec a}(F_I)$ and  define 
$C(F_2, F_1)$ to be the infinite dimensional matrix
\be
C(F_2, F_1) =
\left(
\begin{array}{cc}
\alpha_{F_2,F_1}(\vec{k},\vec{l}) & \beta_{F_2,F_1}(\vec{k},\vec{l})\\           
(\beta_{F_2,F_1}(\vec{k},\vec{l}))^* & (\alpha_{F_2,F_1}(\vec{k},\vec{l}))^*
\end{array}
\right).
\label{cf2f1}
\ee
Then equation (\ref{af1f2}) can be written as 
\be
{\vec a}(F_2)=C(F_2, F_1){\vec a}(F_1)
\label{vecnotation}
\ee
and it follows that 
\be
C(F_3, F_1)=C(F_3, F_2)C(F_2, F_1),  \;\;\; C(F,F) = \bf {1},
\label{cmult} 
\ee
where, as in (\ref{cmultqm}), $\bf 1$ denotes the identity operator.

We can now define the Schrodinger picture as follows.

\subsection{The Representation for ${\hat \phi}(x), {\hat \pi}(x)$} 
Denote ${\hat a}_S (\vec{k}), {\hat a}_S^{\dagger} (\vec{k})$ by $\vec{\hat a}_S$. Given the 
slice $F^A(x)$, define the operators ${\hat b}_F (\vec{k}), {\hat b}_F^{\dagger} (\vec{k})=:\vec{\hat b}(F) $ 
through $\vec{\hat b}(F) = C(F_0, F)\vec{\hat a}_S$ (this is the counterpart of (\ref{defbqm})). Then we have,
analogous to equation (\ref{abqm}), that 
\be
\vec{\hat a}_S = C(F, F_0) \vec{\hat b}(F).
\label{a=cb}
\ee
Define the Fock space ${\cal H}_F$ as the one for which 
$({\hat b}_F (\vec{k}), {\hat b}_F^{\dagger} (\vec{k}))$ are annihilation and creation operators.
Denote states in ${\cal H}_F$ by $\arrowvert \psi,F\rangle$ and the `vacuum' state 
by $\arrowvert 0,F\rangle$ so that ${\hat b}_F (\vec{k})\arrowvert 0,F\rangle =0$ for every $\vec k$.

Consider the (non- seperable) kinematic Hilbert space ${\cal H}_{kin}$ defined as
\be
{\cal H}_{kin} = \bigoplus_F \arrowvert F\rangle \otimes{\cal H}_F,
\label{defhkin}
\ee
where $\arrowvert F\rangle$ carries the representation of the embedding operators as discussed in 
section 3. The linear sum of vector spaces is over all spacelike embeddings $F^A(x)$. 
\footnote{This is akin to the 
sum over the uncountable label set of graphs in LQG \cite{alm2t}. Just as the set of spin network states
organise themselves into a seperable set of states associated with each graph label, here we have
a seperable Fock space associated with each embedding label.\label{f3}}
The inner product on ${\cal H}_{kin}$ is defined in the obvious way from the inner products on the 
embedding sector and on ${\cal H}_F$. To this end consider the linear subspace of ${\cal H}_{kin}$ denoted by 
${\cal D}$ and defined as
\be
{\cal D} =\{\arrowvert \psi\rangle : \arrowvert \psi\rangle = \sum_{I=1}^Na_I\arrowvert F_I\rangle \otimes
                                             \arrowvert \psi_I, F_I\rangle \},     
\label{defd}
\ee
for all finite $N$  and all choices of  complex coefficients $a_I$.
The inner product between $\arrowvert\psi_1\rangle, \arrowvert\psi_2\rangle \in {\cal D}$ with 
$\arrowvert\psi_i\rangle = \sum_{I=1}^Na_{iI}\arrowvert F_I\rangle \otimes
                                             \arrowvert \psi_{iI}, F_I\rangle$, $i=1,2$ is
\be
(\arrowvert\psi_1\rangle, \arrowvert \psi_2\rangle)
=\sum_{I=1}^N a^*_{1I}a_{2I}\langle\psi_{1I}F_I \arrowvert \psi_{2I}F_I\rangle .
\label{hkinip}
\ee
$\cal D$ is dense in ${\cal H}_{kin}$, so that the latter can be obtained by the completion of the former in the 
above inner product.

On any basis element of the form 
$\arrowvert F\rangle \otimes\arrowvert \psi , F\rangle$, the operators 
${\hat X}^A(x),{\hat a}_S (\vec{k}),{\hat a}_S^{\dagger} (\vec{k})$ are defined through
\be
{\hat X}^A(x)\arrowvert F\rangle \otimes\arrowvert \psi , F\rangle
 =F^A(x) \arrowvert F\rangle \otimes\arrowvert \psi , F\rangle  ,                                         
\label{xhatkin}
\ee
\be
{\hat b}_F (\vec{k})\arrowvert F\rangle \otimes\arrowvert \psi , F\rangle
=\arrowvert F\rangle \otimes {\hat b}_F (\vec{k})\arrowvert \psi , F\rangle,
\label{bhatkin}
\ee
\be
{\hat b}^{\dagger}_F (\vec{k})\arrowvert F\rangle \otimes\arrowvert \psi , F\rangle
=\arrowvert F\rangle \otimes {\hat b}^{\dagger}_F (\vec{k})\arrowvert \psi , F\rangle,
\label{bhatadjkin}
\ee
with ${\hat a}_S(\vec{k}),{\hat a}_S^{\dagger} (\vec{k})$ defined in terms of 
${\hat b}_F(\vec{k}),{\hat b}_F^{\dagger} (\vec{k})$ through equation (\ref{a=cb}).
\footnote{Here $\arrowvert \psi , F\rangle \in {\cal H}_F$. We have glossed over the fact that 
${\hat b}_F({\vec k}),{\hat b}^{\dagger}_F({\vec k})$ are operator valued distributions when 
the Cauchy slices are non- compact. We have also glossed over the fact that even when 
appropriately smeared, ${\hat b}_F({\vec k}),{\hat b}^{\dagger}_F({\vec k})$ are not bounded operators
and hence only densely defined. We expect that these technicalities can easily be taken care of;
in the interests of pedagogy, we refrain from a careful treatment of these mathematical niceities.\label{f4}}

It can easily be  verified, with these definitions, that  ${\hat X}^A(x)$ commutes with 
${\hat a}_S (\vec{k}),{\hat a}_S^{\dagger} (\vec{k}),{\hat X}^B(y)$ and that 
$({\hat X}^A(x))^{\dagger}={\hat X}^A(x)$. 

Note that  since   ${\hat b}_F(\vec{k}),{\hat b}_F^{\dagger} (\vec{k})$ are annihilation and creation operators on the 
Fock space ${\cal H}_F$, they are adjoint to each other and have canonical commutation relations on ${\cal H}_F$. 
Note also that ${\hat a}_S (\vec{k}),{\hat a}_S^{\dagger} (\vec{k})$ map states in the subspace 
$\arrowvert F\rangle \otimes {\cal H}_F \subset {\cal H}_{kin}$ into states in the same subspace.
These facts, in conjunction with equation (\ref{a=cb}), the definition of the embedding sector inner product (\ref{embedip}),
and  
the fact that $C(F, F_0)$ is a Bogoliubov transformation, imply that 
${\hat a}_S (\vec{k}),{\hat a}_S^{\dagger} (\vec{k})$  are adjoint to each other on ${\cal H}_{kin}$ 
and have the right commutation relations with each other.

From equation (\ref{a=cb}), the operators ${\vec {\hat a}}_S$ and ${\vec {\hat b}}(F)$ are related by the 
Bogoliubov transformation $C(F, F_0)$. 
The TV results imply that this  Bogoliubov transformation is not unitarily implementable 
on ${\cal H}_{F}$ for generic choices of $F^A(x)$. Due to this fact, we 
are unsure if 
${\hat a}_S({\vec k}),{\hat a}^{\dagger}_S({\vec k})$ are well  defined  operators (after appropriate smearing (see Footnote \ref{f4}))
on (a dense set in) ${\cal H}_{kin}$. However, the formal `working' definition of ${\vec {\hat a}}_S$ 
adopted above through equations (\ref{a=cb}) and (\ref{bhatkin})- (\ref{bhatadjkin}),
and the consequent satisfaction, at a formal level, of the reality conditions and commutation relations involving ${\vec {\hat a}}_S$,
filters down to the rigorous definition of  Dirac observables in section 6C which satisfy the correct commutation
and adjointness relations on (a dense set in) ${\cal H}_{kin}$.
\footnote{We thank Guillermo Mena for discussions on this point.}

To summarise, the representation for ${\hat X}^A(x), {\hat a}_S(\vec{k}),{\hat a}_S^{\dagger} (\vec{k})$ on 
${\cal H}_{kin}$ is such that 
\be
 [{\hat X}^A(x), {\hat X}^B(y)]=0,\;\;\;          [{\hat X}^A(x), \vec{\hat a}_S] =0,
\label{xcom}
\ee
\be
[{\hat a}_S (\vec{k}),{\hat a}_S (\vec{l})]=0, \;\;\; [{\hat a}_S (\vec{k}),{\hat a}_S^{\dagger} (\vec{l})]= \delta (\vec{k},\vec{l}),
\label{acom}
\ee
\be
({\hat X}^A(x))^{\dagger}={\hat X}^A(x),\; \;\; ({\hat a}_S (\vec{k}))^{\dagger}={\hat a}_S^{\dagger} (\vec{k}).
\label{axreality}
\ee
We have deliberately refrained from defining ${\hat H}_G$ to act in the obvious manner because there are problems with such an action.
We now turn to a discussion (and resolution) of these problems.

\section{Problems with ${\hat H}_G$ and their resolution in terms of a treatment of the quantum constraints}
\subsection{Problems with ${\hat H}_G$}
We would like to define ${\hat H}_G$ in the obvious manner by 
${\hat H}_G \arrowvert F\rangle \otimes\arrowvert \psi , F\rangle
= \arrowvert F-G\rangle \otimes\arrowvert \psi , F\rangle$, but this has two problems.
First, assuming $F^A(x)- G^A(x)$ is an embedding, 
if the representation of the matter operators on ${\cal H}_F$ is not unitarily  equivalent to that on ${\cal H}_{F-G}$,
the above state does not lie in ${\cal H}_{kin}$. Second, if $F^A(x)$ is an embedding, $F^A(x)- G^A(x)$ need not be an 
embedding  because embeddings need to satisfy a non- degeneracy condition (described in section 2B; see equation (\ref{nondeg})
and the subsequent discussion)  which ensures that they define a spatial slice in the spacetime $M$.

The first problem can be fixed by enlarging the Hilbert space to 
${\cal H}_X \otimes \oplus_F {\cal H}_F$ (recall that ${\cal H}_X$ is defined in section 3B as the Hilbert space of 
the embedding sector). However the second problem is more acute and is a reflection of a similar problem in the 
quantization of configuration spaces which are not vector spaces, an important example being that of the space
of all Riemannian metrics \cite{ishamnondeg}.

We avoid these problems by considering an alternative set of functions which we will promote to quantum operators.
Thus, instead of choosing our basic Poisson algebra to be generated by the $H_G, X^A(x), \pi (x), \phi (x)$, 
we consider the alternative set of functions $C_A(x),  X^A(x), \pi (x), \phi (x)$, where 
$C_A(x)$ are the constraints of PFT given by equation (\ref{ca}).
From our previous work \cite{tv2}, we anticipate (for $n>1$) that there are obstacles to define 
${\hat C}_A(x)$ as operators. Therefore, motivated by the treatment of the 
quantum (spatial diffeomorphism) constraints in LQG \cite{alm2t}, we adopt the following strategy which focuses on the finite 
canonical transformations generated by $C_A(x)$ rather than on $C_A(x)$ itself.

First we consider the smeared constraints $C(\xi)$ defined by equation (\ref{defcxi}). From the discussion 
in section 2B (see equations (\ref{cxiX}),(\ref{cxiphi}) and (\ref{cxipi})),  we expect that every {\em finite} canonical 
transformation generated by $C(\xi)$ can be labelled by a corresponding spacetime
diffeomorphism $d:M\rightarrow M$ where $d$ is generated by
the vector field $\xi^A$. 
The expected correspondence is as follows.
$d$ maps the point $p$ in $M$ with inertial coordinates $X^A$ to the point $d(p)$.
Denote the inertial coordinates of $d(p)$ by $X^A_d$ and denote the image of the spatial slice $X^A(x)$ 
by  $d$ as 
$X^A_d(x)$.
Then the action of the canonical transformation labelled by $d$ is to map the embedding variables $X^A(x)$ to 
$X^A_d(x)$ and evolve the scalar field data
$(\phi (x), \pi (x))$ to the data $(\phi_d (x), \pi_d (x))$, 
where $(\phi_d (x), \pi_d (x))$ are obtained from
$(\phi (x), \pi (x))$ as follows. Consider the slice $X^A(x)$ in $M$. Let initial data  on this slice be 
$(\phi (x), \pi (x))$. Evolve this data from the slice $X^A(x)$ to the slice $X^A_d(x)$ via the free scalar field evolution equations 
on the flat spacetime $M$. This  evolved data on the slice $X^A_d(x)$ is $(\phi_d (x), \pi_d (x))$.

Note that the data  $(X^A_d(x), \phi_d (x), \pi_d (x))$ (partially) specify a point in phase space
only if $X^A_d(x)$ is also a spacelike slice.
Thus, in order to preserve the phase space, we would like to restrict attention
to those vector fields $\xi^A(X)$ which generate diffeomorphisms which preserve the spacelike property of the 
embeddings $X^A(x)$. 
Let the set of spacelike embeddings be ${\cal E}$. Denote by ${\cal G}({\cal E})$, the subspace of  diffeomorphisms $ d:M\rightarrow M$
such that $X_d^A(x) \in {\cal E}\;\forall X^A(x)\in {\cal E}$.  
It is not clear to us if
there exist any elements
of ${\cal G}({\cal E})$ which are not conformal isometries of the spacetime $(M, \eta_{AB})$.
\footnote{We thank Charles Torre for discussions on this point.}
 The discussion in \cite{karelisham} indicates
that the issue is not trivial. We return to it at the end of this section. For the moment we make the following assumption:\\
{\em ${\cal G}({\cal E})$ exists as an infinite dimensional subgroup}
\footnote{We expect that the vanishing of the constraints $C_A(x)$ is equivalent to the vanishing of the 
smeared constraints $C(\xi )$ provided that the latter holds for sufficiently many $\xi^A(X)$. Since `sufficiently many' 
evidently includes `infinitely many', a minimal requirement on ${\cal G}({\cal E})$ is that it be infinite dimensional.} 
{\em of the group of diffeomorphisms and 
every $d \in{\cal G}({\cal E})$ is generated by some vector field $\xi^A(X)$.}

Consider the canonical 
transformation on phase space labelled by the diffeomorphism $d \in {\cal G}({\cal E})$. We denote the 
operator which implements this transformation in quantum theory by ${\hat U}_d$. 
Then, in addition to the canonical commutation relations between
${\hat \phi}(x), {\hat \pi}(x), {\hat X}^A(x)$, we also need to represent ${\hat U}_d$ such that 
\bea
{\hat U}_{d_2}{\hat U}_{d_1}&=& {\hat U}_{d_1\circ d_2} ,
\label{grp}
\\
{\hat U}_{d^{-1}}&=& {\hat U}_d^{\dagger},
\label{unitary}
\\
{\hat U}_d{\hat X}^A(x){\hat U}_d^{\dagger}&=& {\hat X}^A_d(x),
\label{ux}
\\
{\hat U}_d{\hat \phi}(x){\hat U}_d^{\dagger}&=& {\hat \phi}_d(x),
\label{uphi}
\\
{\hat U}_d{\hat \pi}(x){\hat U}_d^{\dagger}&=& {\hat \pi}_d(x).
\label{upi}
\eea
Several comments are in order. The fact that the equation (\ref{grp}) defines a representation of ${\cal G}({\cal E})$
by right multiplication
follows from equation (\ref{cxipb}). Equation (\ref{unitary}) follows from the fact that the constraints are real.
The right hand side of equation (\ref{ux}) is to be understood as follows. The action of $d$ on the embedding $X^A(x)$ yields the 
 $d$- dependent functionals (one for every value of $A$), $f^A_d[X;x)$, of the embedding. 
Here, in a notation introduced by Kucha{\v r} (see, for example, \cite{karelcyl}),
the left square bracket denotes the fact that for each value of  $A$,  
$f^A$ is a functional of the embedding  and the  round bracket denotes the fact that 
$f^A$ is a function of $x$. Using our notation in the discussion above, we have 
$X^A_d(x) = f^A_d[X;x)$. Then by ${\hat X}^A_d(x)$ we mean that the embedding should be replaced by its corresponding 
quantum operator in  $f^A_d[X;x)$ i.e.${\hat X}^A_d(x):=f^A_d[{\hat X};x)$.
Finally, the right hand sides of equations (\ref{uphi}) and (\ref{upi}) are to be understood in a similar manner.
The classical fields $(\phi_d (x), \pi_d (x))$ are $d$- dependent functionals of the embeddings and the 
fields $(\phi (x), \pi (x))$. Since the underlying dynamics is that of a free field, it follows that the dependence on 
 $(\phi (x), \pi (x))$ is linear. The right hand sides of the equations (\ref{uphi}) and (\ref{upi}) are obtained by substituting 
 the embeddings and the matter fields by the corresponding quantum operators. Since the right hand sides of
(\ref{ux}),(\ref{uphi}) and
(\ref{upi}) are either independent of or linear in the matter fields, and independent of the embedding momenta operators,
there are no operator ordering ambiguities in their definition.

Let $d$ be generated by the vector field $\xi^A(X)$. 
Suppose that ${\hat U}_d$ has the  appropriate continuity in $d$ so that its generator ${\hat C}(\xi)$ can be defined via the action of
${\hat U}_d$ for infinitesmal diffeomorphisms $d$. Then it follows that equation (\ref{grp}) implies that ${\hat C}(\xi)$
satisfies the commutation relations implied by (\ref{cxipb}), that equation (\ref{unitary}) implies that ${\hat C}(\xi)$
is self adjoint, that equations (\ref{ux}), (\ref{uphi}) and (\ref{upi})  imply  that ${\hat C}(\xi)$ satisfies the commutation relations
implied by equations (\ref{cxiX}),(\ref{cxiphi}) and (\ref{cxipi}). It is in this sense that the 
equations (\ref{grp})- (\ref{upi})  provide a  representation of the relevant Poisson bracket algebra and `reality' conditions.

The above discussion was predicated on the assumed existence and properties of ${\cal G}({\cal E})$. It is beyond the scope of this 
paper to analyse the validity of this assumption. From the discussion in \cite{karelisham} it seems unlikely that
${\cal G}({\cal E})$ has the structure of an infinite dimensional group (although the situation of 1+1 dimensions is probably an exception
due to the existence of the infinite dimensional group of conformal isometries).
It turns out that in order to apply the technique of Group Averaging, we need the label `$d$' to take values in some group
which acts on ${\cal E}$. In this regard, we note that the set of bijective maps from ${\cal E}$ to itself has the structure of 
a group (recall that the set of bijective maps of any set defines the Symmetric Group of that set \cite{grptheory}). We denote this
group by ${\cal S}({\cal E})$.

As we shall see, the analysis in the rest of this work holds if
we replace the (putative) group ${\cal G}({\cal E})$ by 
${\cal S}({\cal E})$.  Denote the action of $d\in {\cal S}({\cal E})$ on $X^A(x)\in {\cal E}$ by $X^A_d(x)$.
Let $d_1\circ d_2 \in {\cal S}({\cal E})$ denote the map obtained 
by the composition of the map $d_2 \in {\cal S}({\cal E})$  with $d_1\in{\cal S}({\cal E})$ i.e.
$d_1\circ d_2 (X^A(x)) = d_1 (X^A_{d_2}(x))=: X^A_{d_1\circ d_2}(x)$. As before, define $(\phi_d (x), \pi_d (x))$ as the 
free field evolution of $(\phi (x), \pi (x))$ from  $X^A(x)$ to $X^A_d(x)$.
If ${\cal G}({\cal E})$ with the assumed properties does not exist,
the discussion in this section upto this point serves merely to motivate the 
imposition of equations (\ref{grp})- (\ref{upi}),  with $d$ now taking values in ${\cal S}({\cal E})$.

To summarize,
we replace the  set of quantum operators ${\hat H}_G, {\hat X}^A(x), {\hat \phi} (x) , {\hat \pi}(x)$ by the set
${\hat U}_d, {\hat X}^A(x), {\hat \phi} (x) , {\hat \pi}(x)$.  If our (italicized) assumption on ${\cal G}({\cal E})$ 
is correct then $d$ takes values in
${\cal G}({\cal E})$. If ${\cal G}({\cal E})$ (with the assumed properties) does not exist, 
$d$ takes values in ${\cal S}({\cal E})$. In either case,
we impose the relations (\ref{grp})- (\ref{upi}) in addition to the commutation relations between (and the reality conditions on)
the set ${\hat X}^A(x), {\hat \phi} (x) , {\hat \pi}(x)$. 

{\em The imposition of these conditions, in conjunction with the 
demand that physical states be invariant under the action of ${\hat U}_d$, constitute our definition of the 
Dirac quantization of PFT.}

\subsection{${\cal G}({\cal E})$ versus ${\cal S}({\cal E})$.}

Clearly, the replacement of ${\cal G}({\cal E})$ by ${\cal S}({\cal E})$ is a very non- trivial step.
The group structure of ${\cal S}({\cal E})$ has very little to do with the manifold structure of 
the spacetime $M$. This can be seen through the following example.
Consider two distinct spacelike embeddings $F^A_1(x),F^A_2(x)$. Let the bijective map $d$ be such that (i) it maps
$F^A_1(x)$, $F^A_2(x)$  to the distinct spacelike embeddings  $F^A_{1d}(x)$, $F^A_{2d}(x)$, (ii) it maps 
$F^A_{1d}(x)$ and  $F^A_{2d}(x)$ to $F^A_1(x)$ and $F^A_2(x)$ respectively and (iii) it is the identity on the rest of 
${\cal E}$. We can choose $F^A_1(x)$ and $F^A_2(x)$ to define intersecting  hypersurfaces in $M$.
Since $d$ only needs to be a bijection, we can choose 
$F^A_{1d}(x)$, $F^A_{2d}(x)$ such that they define non- intersecting hypersurfaces in 
$M$. Thus elements of ${\cal S}({\cal E})$ can have a very `discontinuous' action on ${\cal E}$. 
(An even `worse' scenario is if $F^A_1(x)$ and $F^A_2(x)$ are chosen so as to define the same 
hypersurface in $M$ (albeit with different coordinatizations), and 
$F^A_{1d}(x)$, $F^A_{2d}(x)$ are chosen such that they define distinct (say, non- intersecting)
hypersurfaces in $M$.)

Thus, intuitively speaking, ${\cal S}({\cal E})$ corresponds to a huge enlargment of the set of 
gauge transformations generated by the constraints in classical PFT. The justification for using 
${\cal S}({\cal E})$ in the quantization of PFT is an {\sl a posteriori} one- as we shall 
show in section 6C, the resultant quantization of PFT is equivalent to the standard Fock quantization
of the free scalar field on flat spacetime. 
This equivalence relies on the fact that ${\cal S}({\cal E})$ has a transitive action on 
${\cal E}$.

In view of the above discussion, it would be desireable to replace ${\cal S}({\cal E})$ with some  
subgroup thereof which is sensitive to (at least some aspects of) the manifold structure of $M$ and which has 
a transitive action on ${\cal E}$.
\footnote{${\cal G}({\cal E})$ with its assumed properties would be a  candidate if it also had a transitive action;
 however, as mentioned earlier,
we suspect that ${\cal G}({\cal E})$ with the assumed properties does not exist in spacetime dimensions greater than 2.}
This could be attempted by defining a topology on ${\cal E}$
which takes into account properties of $M$ and by restricting attention to continuous bijections (with continuous
inverses) on  ${\cal E}$ which have a transitive action on  ${\cal E}$. We leave such an investigation for 
future work.

\subsection{ Representation of ${\hat U}_d$}

We define the action of ${\hat U}_d$ by its action on  the embedding dependent Fock basis states 
of the subspace $\arrowvert F\rangle \otimes{\cal H}_F \subset {\cal H}_{kin}$ (see equation (\ref{defhkin})). 
For a fixed embedding $F^A(x)$ such a Fock basis consists of the `vacuum'
$\arrowvert F\rangle \otimes\arrowvert 0 , F\rangle $ and the `$N$- particle'  states,
$\arrowvert F\rangle \otimes\prod_{i}^m(b_{F}^{\dagger}(\vec{k_i}))^{n_i}\arrowvert 0 , F\rangle$. Here $N= \sum_{i=1}^m n_i$, and 
the $N$- particle states contain $n_i$ excitations  of momentum $\vec{k_i}$, $i=1,..,m$.
The action of ${\hat U}_d$ is defined as 
\be
{\hat U}_d \arrowvert F\rangle \otimes\arrowvert 0 , F\rangle
= \arrowvert  F_{d^{-1}}\rangle \otimes\arrowvert 0 , F_{d^{-1}}\rangle,
\label{ud0}
\ee
\be
{\hat U}_d \arrowvert F\rangle \otimes  \prod_{i}^m(b_{F}^{\dagger})^{n_i}\arrowvert 0 , F\rangle
= \arrowvert  F_{d^{-1}}\rangle \otimes\prod_{i}^m(b_{F_{d^{-1}}}^{\dagger})^{n_i}\arrowvert 0 , F_{d^{-1}}\rangle.
\label{udbn}
\ee
The action of 
${\hat U}_d$ can be extended to any state in $\arrowvert F\rangle \otimes{\cal H}_F$ (and thence to any state in
${\cal H}_{kin}$)  by linearity.

Next, we show that this action satisfies the equations (\ref{grp})- (\ref{upi}). It suffices to check the action of 
${\hat U}_d$ on the `vacuum' and $N$- particle basis states defined above.

\vspace{2mm}

\noindent {\bf Verification of Equation (\ref{grp})}:
Recall that the notation `$F_{d^{-1}}$' in equation (\ref{udpsi}) signifies the embedding $F_{d^{-1}}$, obtained by the action of 
$d^{-1}$ on the embedding $F^A(x)$.  
Consider the action of 
$d_1$ on ${\cal E}$,
followed by the action of 
$d_2$ 
on ${\cal E}$.
As in equation (\ref{grp}), we denote 
the resultant map on ${\cal E}$
by $d_2 \circ d_1$. Under this map
$X^A$ is mapped to
$X^A_{d_1}$ and thence to
$X^A_{d_2\circ d_1}$.
Thus, we have that 
\bea
{\hat U}_{d_2}{\hat U}_{d_1} \arrowvert F\rangle \otimes\prod_{i}^m(b_{F}^{\dagger})^{n_i}\arrowvert 0 , F\rangle
&=& {\hat U}_{d_2}\arrowvert  F_{{d_1}^{-1}}\rangle \otimes
\prod_{i}^m(b_{F_{{d_1}^{-1}}}^{\dagger})^{n_i}\arrowvert 0 , F_{{d_1}^{-1}}\rangle 
\label{grpveriffirst}
 \\
&=& \arrowvert  F_{{d_2}^{-1}\circ{d_1}^{-1}}\rangle \otimes
\prod_{i}^m(b_{F_{{d_2}^{-1}\circ{d_1}^{-1}}}^{\dagger})^{n_i}\arrowvert 0 , F_{{d_2}^{-1}\circ{d_1}^{-1}}\rangle
\\
&=& \arrowvert F_{({d_1\circ d_2})^{-1}}\rangle \otimes
\prod_{i}^m(b_{F_{({d_1\circ d_2}^{-1})}}^{\dagger})^{n_i}\arrowvert 0 , F_{({d_1\circ d_2})^{-1}}\rangle
\\
&=& {\hat U}_{d_1\circ d_2}\arrowvert F\rangle \otimes \prod_{i}^m(b_{F}^{\dagger})^{n_i}\arrowvert 0 , F\rangle .
\label{grpveriflast}
\eea
Similar considerations hold when we replace the $N$- particle state by the vacuum state $\arrowvert 0 , F\rangle$
in the above equations. This completes the verification of (\ref{grp}).

While the equations (\ref{grpveriffirst})-(\ref{grpveriflast}) are straightforward to check, the notation is a bit cumbersome.
It is more convenient to the use the following notation instead.
Given a state $\arrowvert \psi, F \rangle\in {\cal H}_F$,  we denote the action of ${\hat U}_d$ 
on the state 
$\arrowvert F\rangle \otimes\arrowvert \psi, F\rangle $ (obtained by the expansion of $\arrowvert \psi, F\rangle \in {\cal H}_F$ in the 
Fock basis and the employment of  equations (\ref{ud0}) and (\ref{udbn}) on the basis states)
by 
\be
{\hat U}_d \arrowvert F\rangle \otimes\arrowvert \psi , F\rangle
=: \arrowvert  F_{d^{-1}}\rangle \otimes\arrowvert \psi , F_{d^{-1}}\rangle .
\label{udpsi}
\ee

In this notation we can directly verify (\ref{grp}) for any state $\arrowvert F\rangle \otimes\arrowvert \psi , F\rangle$
as follows:
\bea
{\hat U}_{d_2}{\hat U}_{d_1} \arrowvert F\rangle \otimes\arrowvert \psi , F\rangle
&=& {\hat U}_{d_2}\arrowvert  F_{{d_1}^{-1}}\rangle \otimes
\arrowvert \psi , F_{{d_1}^{-1}}\rangle 
 \\
&=& \arrowvert  F_{{d_2}^{-1}\circ{d_1}^{-1}}\rangle \otimes
\arrowvert \psi , F_{{d_2}^{-1}\circ{d_1}^{-1}}\rangle
\\
&=& \arrowvert F_{({d_1\circ d_2})^{-1}}\rangle \otimes
\arrowvert \psi, F_{({d_1\circ d_2})^{-1}}\rangle
\\
&=& {\hat U}_{d_1\circ d_2}\arrowvert F\rangle \otimes 
\arrowvert \psi , F\rangle .
\eea

\vspace{2mm}

\noindent{\bf Verification of Equation (\ref{unitary})}:

Consider the states $\arrowvert F_1\rangle \otimes\arrowvert \psi_1 , F_1\rangle$,
$\arrowvert F_2\rangle \otimes\arrowvert \psi_2 , F_2\rangle$ where 
$\arrowvert\psi_1, F_1\rangle\in {\cal H}_{F_1}$ and 
$\arrowvert\psi_2, F_2\rangle\in {\cal H}_{F_2}$ and $F_1^A(x), F_2^A(x)$ are embeddings.
Recall that $(,)$ denotes the inner product on ${\cal H}_{kin}$ (see (\ref{hkinip})).
For any 
$d$ 
we have that 
\bea
&({\hat U}_d \arrowvert F_1\rangle \otimes\arrowvert \psi_1 , F_1\rangle,
{\hat U}_d \arrowvert F_2\rangle \otimes\arrowvert \psi_2 , F_2\rangle )&
\nonumber \\
&= 
(\arrowvert (F_1)_{d^{-1}}\rangle \otimes\arrowvert \psi_1 , (F_1)_{d^{-1}}\rangle, 
\arrowvert (F_2)_{d^{-1}}\rangle \otimes\arrowvert \psi_2 , (F_2)_{d^{-1}}\rangle). &
\label{psi1psi2}
\eea
Clearly $(F_1)^A_{d^{-1}}(x) = (F_2)^A_{d^{-1}}(x)$ iff 
$F_1^A(x) = F_2^A(x)$. Also, it is
straightforward to see,  from (\ref{ud0}) and  (\ref{udbn}) in conjunction with our notation, equation (\ref{udpsi}), that 
for any pair of states $\arrowvert\psi , F\rangle,\arrowvert \phi , F\rangle \in {\cal H}_F$  and any $d$,
the following identity holds
\be
\langle \psi, F\arrowvert \phi, F\rangle = 
\langle \psi, F_{d^{-1}}\arrowvert \phi, F_{d^{-1}}\rangle .
\label{findepip}
\ee
It then follows from (\ref{psi1psi2}) that 
\bea
&({\hat U}_d \arrowvert F_1\rangle \otimes\arrowvert \psi_1 , F_1\rangle,
{\hat U}_d \arrowvert F_2\rangle \otimes\arrowvert \psi_2 , F_2\rangle )
\nonumber
\\
& = 0 \; {\rm if \; there \; exists \;} A,x\; {\rm such\; that\; }F_1^A(x)\neq F_2^A (x),
\nonumber \\
& =  \langle \psi_1 , F_1\arrowvert \psi_2 , F_2\rangle \;{\rm if \;} F_1^A(x) = F_2^A(x) \forall x,A .
\label{verifunit}
\eea
It follows from the above equation that ${\hat U}_d$ is unitary. This, in conjunction with 
(\ref{grp}) yields (\ref{unitary}).

\vspace{2mm}

\noindent{\bf Verification of Equation (\ref{ux})}:
We have that 
\bea
{\hat U}_d{\hat X}^A(x){\hat U}^{\dagger}_d \arrowvert F\rangle \otimes\arrowvert \psi , F\rangle
&=&{\hat U}_d{\hat X}^A(x)\arrowvert F_{d}\rangle \otimes\arrowvert \psi , F_{d}\rangle
\\
&=&  F_{d}^A(x){\hat U}_d \arrowvert F_{d}\rangle \otimes\arrowvert \psi , F_{d}\rangle
\label{uxverif1}\\
&=&
{\hat X}^A_d(x)\arrowvert F\rangle \otimes\arrowvert \psi , F\rangle.
\label{uxverif}
\eea
We have used (\ref{xhatkin}) and the definition of ${\hat X}^A_d(x)$ (see the discussion following (\ref{upi}))
in the second and third lines. Note that the second and third lines themselves may be used as a precise definition
of ${\hat X}^A_d(x)$, especially for the case when $d\in {\cal S}({\cal E})$.
This completes the verification of equation (\ref{ux}).

\vspace{2mm}

\noindent{\bf Verification of Equation (\ref{uphi}), (\ref{upi})}:
It is more convenient to show that ${\hat U}_d$ generates the correct evolution on 
the operators ${\hat a}_S(\vec{k}),{\hat a}_S^{\dagger}(\vec{k})$  defined via equations
(\ref{defphis}) and (\ref{defpis}) in section 4A.
Recall from equation (\ref{defak}) in section 4C that the classical counterparts of 
$({\hat a}_S(\vec{k}),{\hat a}_S^{\dagger}(\vec{k}))$ are 
 $(a (\vec{k}),a^* (\vec{k}))$.  We shall refer to the classical correspondent of the action of ${\hat U}_d$ 
as `the finite action of the constraints' even when $d\in {\cal S}({\cal E})$.  In this language, we have that
the classical correspondent of the action of ${\hat U}_d$ on $({\hat a}_S(\vec{k}),{\hat a}_S^{\dagger}(\vec{k}))$
is given by the corresponding finite action of the constraints on $(a (\vec{k}),a^* (\vec{k}))$. 
From the discussion in section 5A, it follows that this action corresponds to classical evolution from the slice
$X^A(x)$ to the slice $X_{d}^A(x)$ where $X^A(x)$ is the embedding part of the PFT phase space data.
Thus in the language and notation of free scalar field evolution employed in section 4C, the modes $a (\vec{k}),a^* (\vec{k})$
are to be thought of as data on the slice $X^A(x)$ and hence are denoted by ${\vec a}(X)$.
Then the finite action of the constraints, labelled by $d$, on the data  ${\vec a}(X)$ is to evolve them
to the data ${\vec a}(X_d)$ by scalar field evolution from the slice $X^A(x)$ to the slice $X_{d}^A(x)$.
Thus, in the notation of section 4C (see, for example, equation (\ref{vecnotation})), we have that 
\be
\vec{a}(X_d) = C(X_d, X)\vec{a}(X) .
\label{cxdxa}
\ee
If ${\hat U}_d$ does indeed generate the correct transformations in quantum theory, then it follows that the
quantum correpondent of the above equation is 
\be
{\hat U}_d{\vec{\hat a}}_S{\hat U}^{\dagger}_d= C ({\hat X}_d, {\hat X}) {\vec {\hat a}}_S .
\label{qcxdxa}
\ee
Here, the embedding dependent linear transformation $C ({\hat X}_d, {\hat X})$ is obtained by replacing 
$X^A(x), X^A_d(x)$ in 
$C(X_d, X)$ by ${\hat X}^A(x), {\hat X}^A_d(x)$ where we have defined
${\hat X}^A_d(x)$ in section  5A (see the discussion following equation (\ref{upi})) and through equations
(\ref{uxverif1}) and (\ref{uxverif}).
 We shall now verify equation (\ref{qcxdxa}).

We have that
\be
{\hat U}_d{\vec{\hat a}}_S{\hat U}^{\dagger}_d\arrowvert F\rangle \otimes\arrowvert \psi , F\rangle
= {\hat U}_d{\vec{\hat a}}_S\arrowvert F_d\rangle \otimes\arrowvert \psi , F_d\rangle .
\ee
Using equation (\ref{a=cb}), we have that 
$\vec{\hat a}_S = C(F_d, F_0)\vec{\hat b}(F_d)$. Using this we obtain
\bea
{\hat U}_d{\vec{\hat a}}_S{\hat U}^{\dagger}_d\arrowvert F\rangle \otimes\arrowvert \psi , F\rangle 
&=& 
 C(F_d, F_0){\hat U}_d\arrowvert F_d\rangle \otimes\vec{\hat b}(F_d)\arrowvert \psi , F_d\rangle
\\
&=&
C(F_d, F_0)\arrowvert F\rangle \otimes \vec{\hat b}(F)\arrowvert \psi , F\rangle
\label{5b1}
\\
&=&
C(F_d, F_0)C(F_0, F){\vec{\hat a}}_S\arrowvert F\rangle \otimes\arrowvert \psi , F\rangle
\label{5b2}
\\
&=&
C(F_d, F){\vec{\hat a}}_S\arrowvert F\rangle \otimes\arrowvert \psi , F\rangle
\label{5b3}
\\
&=&
C({\hat X}_d, {\hat X}){\vec{\hat a}}_S\arrowvert F\rangle \otimes\arrowvert \psi , F\rangle .
\label{5b4}
\eea

Here we have used (\ref{a=cb}) in  (\ref{5b2}) and (\ref{cmult}) in (\ref{5b3}).
We have also used the easily verifiable fact (see the remarks at the end of section 4C) that ${\hat X}^A(x)$ commutes with 
${\vec {\hat a}}_S$. 
This completes our discussion of the equations (\ref{uphi}) and (\ref{upi}).


To summarise, we have verified that ${\cal H}_{kin}$ provides a Hilbert space representation for the 
equations (\ref{xcom})- (\ref{axreality}), (\ref{grp})- (\ref{ux}) and (\ref{qcxdxa}).
This implies that we have  defined a $*$- representation for the set
of operators 
$({\hat U}_d, {\hat X}^A(x), {\hat a}_S({\vec{k}), {\hat a}^{\dagger}_S ({\vec k})})$ 
(or, equivalently,  using equations  (\ref{defphis}),(\ref{defpis}), 
of the set $({\hat U}_d, {\hat X}^A(x), {\hat\phi}(x), {\hat \pi (x)})$).
We now proceed to an implementation of Dirac quantization along the lines employed in LQG \cite{alm2t}.

\section{Dirac quantization}

Just as in the case of  the spatial diffeomorphism constraints in LQG, ${\hat U}_d$ does not have the appropriate 
continuity in $d$ to allow a definition of its (putative) generator ${\hat C}_A(x)$. Clearly, this is so for
$d\in {\cal G}({\cal E})$ and is expected to be so for $d\in {\cal S}({\cal E})$ for any reasonable topology 
on ${\cal S}({\cal E})$. If ${\hat C}_A(x)$ is not defined, the identification of its kernel as the space of 
physical states is not possible. In such a case we can still 
use the LQG method
of Group Averaging \cite{dongrpavg,alm2t} to construct the physical state space.  
After some preliminary remarks in section 6A,
we discuss the construction of the physical Hilbert space ${\cal H}_{phys}$
in 
section 6B and the action of Dirac observables on physical states in section 6C. We shall see that the resultant quantization
is 
equivalent to the standard Fock space quantization. In section 6D we show how to obtain the action of 
${\hat C}_A(x)$ on physical states as  functional Schrodinger equations in the case when operator evolution is unitary and
all the embedding dependent Hilbert spaces $\{{\cal H}_F\}$ are identical.
We shall attempt a reasonably self contained exposition. However the reader is urged to consult the 
references \cite{alm2t,dongrpavg} for details regarding the Group Averaging technique. We shall intersperse our 
exposition with remarks pertaining to the analogy with structures in LQG. These remarks may be ignored by 
readers unfamiliar with LQG.

\subsection{Preliminary Remarks.}

\noindent
{\bf (i)}
Given $F_1^A(x), F_2^A(x)\in {\cal E}$, ${\cal H}_{F_1}$ can be identified with ${\cal H}_{F_2}$ by an identification of
the Fock basis states as follows. Identify $\arrowvert 0 , F_1\rangle$ with 
$\arrowvert 0 , F_2\rangle$ and identify the $N$ particle states (see section 5C) 
$\prod_{i}^m(b_{F_1}^{\dagger}(\vec{k_i}))^{n_i}\arrowvert 0 , F_1\rangle$
with $\prod_{i}^m(b_{F_2}^{\dagger}(\vec{k_i}))^{n_i}\arrowvert 0 , F_2\rangle$. Since any state can be expanded in the above Fock 
basis, 
every state in ${\cal H}_{F_1}$ has a counterpart in ${\cal H}_{F_2}$ and vice versa. We shall use the notation
$\arrowvert \psi , F_1\rangle$ and $\arrowvert \psi , F_2\rangle$ to denote the states so identified. Clearly, if 
$F_1^A(x) = F_{2d}^A(x)$ for some (not necessarily unique) choice of $d$, we have that 
\be
\arrowvert F_2\rangle \otimes\arrowvert \psi , F_2\rangle := {\hat U}_d\arrowvert F_1\rangle \otimes\arrowvert \psi , F_1\rangle ,
\label{psif}
\ee
which is consistent with the notation of equation (\ref{udpsi}).

\vspace{2mm}

\noindent
{\bf (ii)} Note that even if $d\in {\cal G}({\cal E})$ , $d$ defines a bijective map on ${\cal E}$. This follows from the fact that
$\forall F^A(x)\in {\cal E}$, there exists ${\bar F}^A(x)=F^A_{d^{-1}}(x)  \in {\cal E}$ so that 
$F^A(x) = {\bar F}_d^A(x)$.

\vspace{2mm}

\noindent
{\bf (iii)} Let ${\cal D}^*$ denote the vector space of complex linear mappings from ${\cal D} \subset {\cal H}_{kin}$ (see equation
(\ref{defd})) to the set of complex numbers. 
${\cal D}^*$ is called the {\em algebraic dual} to ${\cal D}$. Elements of ${\cal D}^*$ are called {\em distributions}.
Consider any operator ${\hat A}$ on ${\cal H}_{kin}$ such that ${\hat A}^{\dagger}$ maps ${\cal D}$ into itself.
Define the action of ${\hat A}$ on  ${\cal D}^*$ as follows. Let ${\bf \Psi} \in {\cal D}^*$ map $ \arrowvert \phi \rangle \in {\cal D}$
to the complex number ${\bf \Psi}(\arrowvert \phi \rangle )$. Define 
${\hat A}{\bf \Psi}$ by 
\be
{\hat A}{\bf \Psi}(\arrowvert \phi \rangle ):=
{\bf \Psi}({\hat A}^{\dagger}\arrowvert \phi \rangle ) .
\label{dualaction}
\ee
Let ${\hat A}, {\hat B}$ be operators such that the operators 
${\hat A}^{\dagger},{\hat B}^{\dagger},{\hat A}^{\dagger}{\hat B}^{\dagger},{\hat B}^{\dagger}{\hat A}^{\dagger}$
map ${\cal D}$ into itself. Then we have that 
\be
{\hat A}{\hat B}{\bf \Psi}(\arrowvert \phi \rangle )
= {\hat B}{\bf \Psi}({\hat A}^{\dagger}\arrowvert \phi \rangle )
={\bf \Psi}({\hat B}^{\dagger}{\hat A}^{\dagger}\arrowvert \phi \rangle ),
\ee
so that the the action of the commutator $[{\hat A},{\hat B}]$ on ${\bf \Psi}$ is 
\be
[{\hat A},{\hat B}]{\bf \Psi}(\arrowvert \phi \rangle )
={\bf \Psi}([{\hat B}^{\dagger},{\hat A}^{\dagger}]\arrowvert \phi \rangle )
={\bf \Psi}([{\hat A},{\hat B}]^{\dagger}\arrowvert \phi \rangle ).
\label{dualrep}
\ee
Equation (\ref{dualrep}) imples that ${\cal D}^*$ provides an anti- representation for the operators 
${\hat A}, {\hat B}$.

\subsection{Physical states through Group Averaging.}

For the remainder of this work, we shall restrict attention to the case that $d\in {\cal S}({\cal E})$. Our considerations
will also apply unchanged if ${\cal G}({\cal E})$ is such that given $F_1^A(x), F_2^A(x) \in {\cal E}$, there exists
$d\in {\cal G}({\cal E})$  such that $F_{1d}^A(x)= F_2^A(x)$. If ${\cal G}({\cal E})$ does not have this property there 
may be superselection sectors in  the physical state space constructed via Group Averaging. Since we do not even know if
${\cal G}({\cal E})$ exists, much less the structure of ${\cal G}({\cal E})$, we leave the analysis for this case 
for future work.

\subsubsection{Construction of the space of physical states} 

Physical states must be invariant under the action of ${\hat U}_d$ i.e. $\Psi \in {\cal H}_{phys}$ iff $\forall d$,
${\hat U}_d\Psi = \Psi$. We can formally construct such a state 
as follows.
Define $\Psi$ by the formal sum
\be
\Psi = 
\sum_{F\in {\cal E}}\arrowvert F\rangle \otimes\arrowvert \psi , F\rangle .
\label{defpsi}
\ee
Here, we have used the notation defined 
in 
{\bf (i)} above.
Then we have that 
\bea
{\hat U}_d\Psi &=& {\hat U}_d \sum_{F\in {\cal E}}\arrowvert F\rangle \otimes\arrowvert \psi , F\rangle 
\\
&=& \sum_{F_{d^{-1}}\in {\cal E}} \arrowvert F_{d^{-1}}\rangle \otimes\arrowvert \psi , F_{d^{-1}}\rangle
\\
&=&
\sum_{F\in {\cal E}}\arrowvert F\rangle \otimes\arrowvert \psi , F\rangle = \Psi
\label{diffinv}
\eea

The formal sum in (\ref{defpsi}), in the language of Group Averaging, corresponds to the choice of 
a discrete measure on ${\cal S}({\cal E})$. To see this, we note that we can define
\be
\Psi = 
(\sum_{d\in {\cal S}({\cal E})} {\hat U}_d )\arrowvert F_1\rangle \otimes\arrowvert \psi , F_1\rangle .
\label{defpsiavg}
\ee
where $F_1^A(x)$ is some fixed embedding and the discrete measure on 
${\cal S}({\cal E})$
is such that the right hand sides of (\ref{defpsiavg}) and (\ref{defpsi}) are identical.
Note that $F_1^A(x)$ has a non- trivial `isotropy group' \cite{alm2t} i.e. the subgroup 
of ${\cal S}({\cal E})$
which leaves $F_1^A(x)$ invariant. The sum in equation (\ref{defpsiavg}) is, therefore, not over the entire 
group ${\cal S}({\cal E})$,
but only over the orbit of  the state $\arrowvert F_1\rangle \otimes\arrowvert \psi , F_1\rangle$.

The sum (\ref{defpsi}) is a formal one. To make our considerations well defined, it is appropriate to 
think of $\Psi$ as defining the distribution ${\bf \Psi} \in {\cal D}^*$ through
\be
{\bf \Psi}
=\sum_{F\in {\cal E}}
\langle\psi , F\arrowvert \otimes\langle  F\arrowvert .
\label{dualpsi}
\ee
The sum in (\ref{dualpsi}) is to be interpreted through the action of ${\bf \Psi}$ on ${\cal D}$ as follows.
${\bf \Psi}$ maps the state $\arrowvert F_1\rangle \otimes\arrowvert \phi , F_1\rangle$
to the complex number ${\bf \Psi}(\arrowvert F_1\rangle \otimes\arrowvert \phi , F_1\rangle )$ where
\bea
{\bf \Psi}(\arrowvert F_1\rangle \otimes\arrowvert \phi , F_1\rangle )
&=& (\sum_{F\in {\cal E}}
\arrowvert F\rangle \otimes\arrowvert \psi , F\rangle ,\arrowvert F_1\rangle \otimes\arrowvert \phi , F_1\rangle )
\\
&=&  \sum_{F\in {\cal E}} \delta_{F,F_1}\langle\psi , F_1\arrowvert\phi , F_1\rangle
\\
&=&
\langle\psi , F_1\arrowvert\phi , F_1\rangle ,
\label{psiphi}
\eea
where we have used the definition (see equation (\ref{hkinip})) of the inner product on ${\cal H}_{kin}$ in the second line.

The action of ${\bf \Psi}$ can be extended to all of ${\cal D}$ by linearity. 
The 
invariance of $\Psi$ 
(\ref{diffinv}) translates to the easily verifiable statement that $\forall \arrowvert \phi \rangle \in {\cal D}$,
\be
{\hat U}_d {\bf \Psi} (\arrowvert \phi \rangle ) ={\bf \Psi}({\hat U}_d^{\dagger}\arrowvert \phi \rangle )=
 {\bf \Psi} (\arrowvert \phi \rangle ).
\label{diffinvdual}
\ee
From (\ref{defpsiavg}) it follows that ${\bf \Psi}$ can also be defined by
\be
{\bf \Psi}=\sum_d
\langle\psi , F_1\arrowvert \otimes\langle  F_1\arrowvert {\hat U}^{\dagger}_d ,
\label{psiavgdual}
\ee
where  the sum over $d$ is again to be interpreted as a sum over the orbit of the state 
$\arrowvert F_1\rangle \otimes\arrowvert \psi , F_1\rangle$ rather than over the entire group ${\cal S}({\cal E})$.
Since the sum in (\ref{defpsiavg}) and (\ref{psiavgdual}) is over  elements of (a (non- unique) subset of)
the group ${\cal S}({\cal E})$,
this technique of constructing physical states from kinematic ones
 is called
Group Averaging. 

Denote the finite linear span of the set of invariant distributions (i.e. those defined through
equation (\ref{dualpsi}) for all choices of $\arrowvert \psi , F\rangle$) by ${\cal D}^*_{phys}$.
Thus ${\bf \Phi} \in {\cal D}^*_{phys}$ iff ${\bf \Phi}=\sum_{I=1}^N c_I{\bf \Phi}_I$, with $c_I$ 
being complex numbers and each ${\bf \Phi}_I$ being of the form (\ref{dualpsi}),
\be
{\bf \Phi}_I = \sum_{F\in {\cal E}}
\langle\phi_I, F\arrowvert \otimes\langle  F\arrowvert .
\label{1}
\ee
We shall see that ${\cal D}^*_{phys}$ can be equipped with a natural inner product thus converting it 
into the Hilbert space ${\cal H}_{phys}$. Before introducing this inner product, it is useful to 
understand the structure of ${\cal D}^*_{phys}$ better. We do this by defining the Group Averaging map
$\eta$ \cite{alm2t}.

\subsubsection{The Group Averaging map $\eta$.}
 
Equation (\ref{psiavgdual}) defines a map from states of the form
$\arrowvert F\rangle \otimes\arrowvert \psi , F\rangle$ to states in ${\cal D}^*_{phys}$. This map can be
extended to all of ${\cal D}$ by anti- linearity. In the notation of Reference \cite{alm2t}, 
this map is denoted by $\eta$. Explicitly, $\eta:{\cal D} \rightarrow {\cal D}^*_{phys}$ is defined as follows. Let 
$\arrowvert \phi\rangle \in {\cal D}$ so that 
\be
\arrowvert \phi\rangle = \sum_{I=1}^Na_I\arrowvert F_I\rangle \otimes
                                             \arrowvert \phi_I, F_I\rangle .
\label{0.9}
\ee 
Then $\eta \arrowvert \phi\rangle =: {\bf \Phi}$ is given  by the following expression:
\be
{\bf \Phi}
= \sum_{I=1}^Na_I^* (\sum_d\langle\phi_I , F_I\arrowvert \otimes\langle  F_I\arrowvert {\hat U}^{\dagger}_d)
= \sum_{I=1}^Na_I^* {\bf \Phi}_I
\label{1.1}
\ee
with ${\bf \Phi}_I= \sum_{F\in {\cal E}}
\langle\phi_I, F\arrowvert \otimes\langle  F\arrowvert$ .

From the definition of ${\cal D}^*_{phys}$, it follows that the anti- linear map $\eta$ is onto.
The map $\eta$ has a non- trivial kernel. This can be seen as follows.
The action of ${\bf \Phi}$ defined in equation (\ref{1.1}) on any state 
$\arrowvert \psi \rangle \in {\cal D}$, 
\be
\arrowvert\psi\rangle := \sum_{J=1}^Mb_J\arrowvert G_J\rangle \otimes
                                             \arrowvert \psi_J, G_J\rangle
\label{1.4}
\ee 
is
\be
{\bf \Phi}(\arrowvert\psi\rangle )
=\sum_{I=1}^N\sum_{J=1}^M a_I^*b_J\langle\phi_I , G_J\arrowvert \psi_J, G_J\rangle .
\label{aibj}
\ee
From equation (\ref{psif}) and the fact that ${\hat U}_d$ is unitary, it follows that 
${\bf \Phi}(\arrowvert\psi\rangle )$ can also be written as 
\be
{\bf \Phi}(\arrowvert\psi\rangle )
=\sum_{I=1}^N\sum_{J=1}^M a_I^*b_J\langle\phi_I , F\arrowvert \psi_J, F\rangle ,
\label{1.5}
\ee
for any $F^A(x) \in {\cal E}$. This implies that ${\bf \Phi}(\arrowvert\psi\rangle )= 0$
 $\forall \arrowvert\psi\rangle \in {\cal D}$  iff $\forall F^A(x) \in {\cal E}$,
\be
\sum_{I=1}^Na_I^* \langle\phi_I , F\arrowvert \otimes\langle  F\arrowvert =0.
\label{2}
\ee
Equation (\ref{2}) implies that the kernel of $\eta$ consists of
all $\arrowvert\phi\rangle \in {\cal D}$, 
$\arrowvert\phi\rangle = \sum_{I=1}^Na_I\arrowvert F_I\rangle \otimes
                                             \arrowvert \phi_I, F_I\rangle \in {\cal D}$
with $a_I, \arrowvert \phi_I, F_I\rangle$ such that equation (\ref{2}) holds.

The above structure of the kernel implies that every element of ${\cal D}^*_{phys}$
is of the form (\ref{dualpsi}). More precisely, let ${\bf \Phi}$ be an arbitrary element of 
${\cal D}^*_{phys}$. Then, as can be verified, there exists a unique state 
$\arrowvert \phi , F\rangle \in {\cal H}_F$ such that $\forall F^A(x) \in {\cal E}$
\be 
{\bf \Phi} = \eta \arrowvert F\rangle \otimes\arrowvert \phi , F\rangle
\label{2.1}
\ee

\subsubsection{The Inner Product on ${\cal D}^*_{phys}$.}

The map $\eta$ endows ${\cal D}^*_{phys}$ with the following inner product.
Let ${\bf \Psi},{\bf \Phi} \in {\cal D}^*_{phys}$. Then their inner product is defined as 
\be
({\bf \Psi},{\bf \Phi} ) = {\bf \Phi} (\arrowvert \psi\rangle ),
\label{2.5}
\ee
where $\arrowvert \psi \rangle \in {\cal D}$ is any state in ${\cal D}$ such that 
$\eta \arrowvert \psi \rangle  = {\bf \Psi}$.  

We note the following:

\vspace{2mm}

\noindent {\bf (a)} The right hand side of equation (\ref{2.5}) is independent of the choice of $\arrowvert \psi\rangle$:
Let $\arrowvert \psi_1\rangle , \arrowvert \psi_2\rangle$ be such that 
$\eta \arrowvert \psi_1\rangle = \eta \arrowvert \psi_2\rangle ={\bf \Psi}$. Then 
$\arrowvert \psi_1\rangle - \arrowvert \psi_2\rangle$ is in the kernel of $\eta$. From equation (\ref{2}) this means that 
$\arrowvert \psi_1\rangle - \arrowvert \psi_2\rangle
=\sum_{J=1}^Mb_J\arrowvert G_J\rangle \otimes
                                             \arrowvert \psi_J, G_J\rangle$ with $b_J, \arrowvert \psi_J, G_J\rangle$
such that 
\be
\sum_{J=1}^Mb_J^* \langle\psi_J , F\arrowvert \otimes\langle  F\arrowvert =0, \;\; \forall F^A(x) \in {\cal E}.
\label{3}
\ee
Let ${\bf \Phi}=\eta \arrowvert \phi\rangle$ with $\arrowvert \phi\rangle\in {\cal D}$ given by 
$\arrowvert \phi\rangle = \sum_{I=1}^Na_I\arrowvert F_I\rangle \otimes
                                             \arrowvert \phi_I, F_I\rangle$. 
Then 
${\bf \Phi}(\arrowvert \psi_1\rangle - \arrowvert \psi_2\rangle)$ is given by equation (\ref{1.5}), the right hand side of which 
vanishes as a result of equation (\ref{3}).
Thus ${\bf \Phi}(\arrowvert \psi_1\rangle ) ={\bf \Phi}(\arrowvert \psi_2\rangle )$.

\vspace{2mm}

\noindent {\bf (b)} The inner product is Hermitian:  
Let ${\bf \Phi} , \arrowvert\psi\rangle$ be given by equations (\ref{1.1}),(\ref{1.4}) with 
$a_I, b_J$ not necessarily satisfying equations (\ref{2}) and (\ref{3}). Then from equation (\ref{1.5}), we have that 
\be
({\bf \Psi},{\bf \Phi} ) =({\bf \Phi} (\arrowvert \psi\rangle ))
=
\sum_{I=1}^N\sum_{J=1}^M a_I^*b_J\langle\phi_I , F\arrowvert \psi_J, F\rangle
= ({\bf \Phi},{\bf \Psi} )^*
\ee

\vspace{2mm}

\noindent {\bf (c)} Equation (\ref{2.5}) is linear in its second element and anti- linear in its first element.
This follows from the linear vector space structure of ${\cal D}^*_{phys}$ and the anti- linearity of
the map $\eta$.

\vspace{2mm}

\noindent {\bf (d)} The inner product is positive definite: Let 
${\bf \Phi}= \eta\arrowvert \phi\rangle $ 
be given by equations (\ref{0.9}) and (\ref{1.1}) with $a_I$ not necessarily satisfying equation (\ref{2}) . Then 
\be
({\bf \Phi},{\bf \Phi} ) ={\bf \Phi} (\arrowvert \phi\rangle )= 
\sum_{I=1}^N\sum_{J=1}^N a_I^*a_J\langle\phi_I , F\arrowvert \phi_J, F\rangle
=\langle\phi , F\arrowvert \phi , F\rangle
\ee
with $\arrowvert \phi , F\rangle = \sum_{I=1}^N a_I\arrowvert \phi_I, F\rangle$.
Thus $({\bf \Phi},{\bf \Phi} ) \geq 0$
and vanishes only if $\arrowvert \phi , F\rangle =0$ i.e. only if
$\arrowvert \phi\rangle$ is in the kernel of $\eta$ (see equation (\ref{2})).

\vspace{2mm} 

\noindent {\bf (e)} As in Footnote \ref{f4}, we shall gloss over mathematical niceities related to 
the distinction between operators and operator valued distributions, as well as issues related
to unbounded operators and dense domains. However, we do note here that 
 the completion of 
the space ${\cal D}^*_{phys}$ in  the inner product (\ref{2.5}) does not enlarge 
${\cal D}^*_{phys}$ so that, as vector spaces, ${\cal D}^*_{phys}= {\cal H}_{phys}$. 
The interested reader may easily verify this fact by using  equation (\ref{2.1}) and noting that  ${\cal H}_F, F^A(x)\in {\cal E}$ are 
Hilbert spaces and hence Cauchy complete.

\subsection{Dirac Observables.}

\subsubsection{Classical Theory.}

From the work of Kucha{\v r} \cite{kareliyer,karel1+1c} it follows that Dirac observables can be identified with the values
of the phase space variables $(\phi (x), \pi (x) )$, or equivalently, $\vec{a} (X)$, (see the discussion before equation 
(\ref{cxdxa}) for the definition of $\vec{a} (X)$ ) at some initial embedding, say $F^A_0(x)$. 
We denote the Dirac observables corresponding to the values of $\vec{a} (X)$ at $F^A_0(x)$ by 
 $\vec{a}_D = a_D(\vec{k}),a_D^*(\vec{k})$. Thus $\vec{a}_D$ is   given by the expression
\be
\vec{a}_D := C(F_0, X) \vec{a} (X). 
\label{defad}
\ee
The following argument shows that $a_D(\vec{k}),a_D^*(\vec{k})$  are indeed Dirac observables.
Consider a  finite transformation generated by the constraints which
evolves the 
data $(X^A(x), \vec{a} (X))$  to 
$(X_d^A(x), \vec{a} (X_d))$.
Denote $\vec{a}_D$ evaluated on the new data by $(\vec{a}_D)_d$. Then we have that 
\bea
(\vec{a}_D)_d &= & C(F_0, X_d) \vec{a} (X_d) \nonumber \\
&=& C(F_0, X_d)C (X_d, X) \vec{a} (X) \nonumber \\
&=&  C(F_0, X) \vec{a} (X) = \vec{a}_D,
\eea
where we used equations (\ref{vecnotation}) and (\ref{cmult}) in the second line.
Thus $a_D(\vec{k}),a_D^*(\vec{k})$ are invariant under the action of the constraints. Hence 
they are classical Dirac observables. 

Clearly, since $\vec{a}_D$ is related to $\vec{a} (X)$ via a canonical transformation only 
dependent on $X^A(x)$, it follows that the Poisson brackets between $a_D(\vec{k}),a_D^*(\vec{k})$ are 
given by 
\be
\{a_D(\vec{k}),a_D(\vec{l})\} =0 =\{a_D^*(\vec{k}),a_D^*(\vec{l})\}, \;\;\;
\{a_D(\vec{k}),a_D^*(\vec{l})\}= i \delta({\vec k}, {\vec l})
\label{3.5}
\ee

\subsubsection{Quantum Theory.}

The quantum correspondent of equation (\ref{defad}) is 
\be
\vec{\hat a}_D := C(F_0, {\hat X}) \vec{\hat a}_S, 
\label{defadq}
\ee
where $C(F_0, {\hat X})$ is defined by replacing $X^A(x)$ by ${\hat X}^A(x)$ in the expression for
$C(F_0, X)$. The action of $\vec{\hat a}_D$ on $\arrowvert F\rangle \otimes\arrowvert \psi , F\rangle$
is given by 
\bea
\vec{\hat a}_D\arrowvert F\rangle \otimes\arrowvert \psi , F\rangle
&=&  C(F_0, {\hat X}) \vec{\hat a}_S\arrowvert F\rangle \otimes\arrowvert \psi , F\rangle
\\
&=&  C(F_0, F)\vec{\hat a}_S\arrowvert F\rangle \otimes\arrowvert \psi , F\rangle
\label{i}
\\
&=& C(F_0, F)C(F, F_0)\arrowvert F\rangle \otimes \vec{\hat b}(F)\arrowvert \psi , F\rangle
\label{ii}
\\
&=& \vec{\hat b}(F)\arrowvert F\rangle \otimes \arrowvert \psi , F\rangle .
\label{iii}
\eea
We have used the fact that ${\hat X}^A(x)$ commutes with $\vec{\hat a}_S$ (see equation (\ref{xcom}) in (\ref{i}).
Equation (\ref{i}) follows from the definition of $\vec{\hat a}_S$ in equation (\ref{a=cb}). 
Equation (\ref{iii}) implies that $\vec{\hat a}_D$  has the following properties.

\vspace{2mm}

\noindent {\bf (a)} $\vec{\hat a}_D$ maps physical states to physical states:
Let ${\bf \Phi} \in {\cal D}^*_{phys}$. Then it follows from equation (\ref{2.1}) that 
${\bf \Phi}$ is of the form 
${\bf \Phi} = \eta \arrowvert\phi\rangle$, $\arrowvert\phi\rangle=\arrowvert F\rangle \otimes\arrowvert \phi , F\rangle$.
Let $\arrowvert\psi\rangle \in {\cal D}$, 
$\arrowvert\psi\rangle= \arrowvert G\rangle \otimes\arrowvert \psi, G\rangle$. Then 
\bea
{\hat a}_D(\vec{k}){\bf \Phi}(\arrowvert\psi\rangle)
&=&
{\bf \Phi}({\hat a}_D^{\dagger}(\vec{k})\arrowvert\psi\rangle)
\\
&=&
{\bf \Phi}(\arrowvert G\rangle \otimes {\hat b}_G^{\dagger}(\vec{k})\arrowvert\psi\rangle)
\\
&=&
\langle\phi , G\arrowvert {\hat b}_G^{\dagger}(\vec{k})\arrowvert \psi , G\rangle
\\
&=:&
{\bf \Phi}_{{\hat a}_D(\vec{k})} (\arrowvert\psi\rangle),
\label{4}
\eea
where 
\be 
{\bf \Phi}_{{\hat a}_D(\vec{k})} = \eta ({\hat b}_F(\vec{k})\arrowvert F\rangle \otimes\arrowvert \phi, F\rangle).
\label{5}
\ee
Since states of the form $\arrowvert G\rangle \otimes\arrowvert \psi, G\rangle$
are a basis for ${\cal D}$, and since $\eta ({\hat b}_F(\vec{k})\arrowvert\phi\rangle)$ is independent of 
the choice of $F^A(x) \in {\cal E}$, it follows that ${\hat a}_D(\vec{k}){\bf \Phi}$ is of the form 
(\ref{dualpsi}) and is hence a physical state. Similar considerations hold for 
${\hat a}_D(\vec{k})^{\dagger}$.

\vspace{2mm}

\noindent {\bf (b)} The action of the operators  ${\hat a}_D(\vec{k}),{\hat a}_D^{\dagger}(\vec{k})$ on physical states 
provides an anti- representation of their Poisson bracket relations 
(\ref{3.5}): 
Given 
${\bf \Phi} \in {\cal D}^*_{phys}$
 and $\arrowvert \psi\rangle=\arrowvert F\rangle \otimes\arrowvert \psi , F\rangle$,
we have that 
\bea
({\hat a}_D(\vec{k}){\hat a}_D^{\dagger}(\vec{l}){\bf \Phi})(\arrowvert \psi\rangle)
&=&
({\hat a}_D^{\dagger}(\vec{l}){\bf \Phi})(\arrowvert F\rangle \otimes {\hat b}_F^{\dagger}(\vec{k})\arrowvert\psi ,F\rangle)
\\
&=&
{\bf \Phi}(\arrowvert F\rangle \otimes{\hat b}_F(\vec{l}) {\hat b}_F^{\dagger}(\vec{k})\arrowvert\psi ,F \rangle),
\label{6.42}
\eea
so that $[{\hat a}_D(\vec{k}){\hat a}_D(\vec{l})^{\dagger}]=
[{\hat b}_F(\vec{l}), {\hat b}_F^{\dagger}(\vec{k})] = \delta(\vec{k},\vec{l})$.
Similar calculations, with the same choice of $\arrowvert \psi\rangle$ as in the above equations
  show that the commutators $[{\hat a}_D(\vec{k}),{\hat a}_D(\vec{l})]$, 
$[{\hat a}^{\dagger}_D(\vec{k}),{\hat a}^{\dagger}_D(\vec{l})]$ are mapped to the 
commutators  $[{\hat b}_F^{\dagger}(\vec{l}),{\hat b}_F^{\dagger}(\vec{k})]$,$[{\hat b}_F(\vec{l}),{\hat b}_F(\vec{k})]$.
Note that (i)${\hat b}_F(\vec{k}),{\hat b}_F^{\dagger}(\vec{k})$  are annhilation, creation operators on ${\cal H}_F$, 
(ii) equation (\ref{6.42}) (and similar equations for the other commutators) is independent of the choice of $F^A(x)$
and (iii) states of the form $\arrowvert F\rangle \otimes\arrowvert \psi , F\rangle, F^A(x) \in {\cal E}$ provide a basis for
${\cal D}$.

It follows from (i) - (iii) that equation (\ref{iii}) defines  an anti- representation, on ${\cal H}_{phys}$,  of the 
Poisson algebra generated by the Poisson brackets  (\ref{3.5}).

\vspace{2mm}

\noindent {\bf (c)} The operators ${\hat a}_D(\vec{k}),{\hat a}_D^{\dagger}(\vec{k})$ are adjoint to each other on 
${\cal H}_{phys}$: 
Let ${\bf \Psi}_i = \eta \arrowvert \psi_i\rangle$, 
$\arrowvert \psi_i\rangle=\arrowvert F\rangle \otimes\arrowvert \psi_i , F\rangle$, $i=1,2$ (Equation (\ref{2.1}) implies that 
this choice of $\arrowvert \psi_i\rangle$ involves no loss of generality).
Then we have that 
\bea
({\bf \Psi}_1,{\hat a}_D(\vec{k}){\bf \Psi}_2))
&=&
{\bf \Psi}_2({\hat b}^{\dagger}_F(\vec{k})\arrowvert F\rangle \otimes\arrowvert \psi_1 , F\rangle)
\\
&=&
\langle\psi_2 , F\arrowvert{\hat b}^{\dagger}_F(\vec{k})\arrowvert \psi_1 , F\rangle,
\\
({\bf \Psi}_2,{\hat a}_D^{\dagger}(\vec{k}){\bf \Psi}_1))
&=&
{\bf \Psi}_1({\hat b}_F(\vec{k})\arrowvert F\rangle \otimes\arrowvert \psi_2 , F\rangle)
\\
&=&
\langle\psi_1 , F\arrowvert {\hat b}_F(\vec{k})\arrowvert \psi_2 , F\rangle .
\eea
Since ${\hat b}_F(\vec{k}), {\hat b}^{\dagger}_F(\vec{k})$ are Hermitian conjugates on ${\cal H}_F$, it follows that
$({\bf \Psi}_1,{\hat a}_D(\vec{k}){\bf \Psi}_2))^*= ({\bf \Psi}_2,{\hat a}_D^{\dagger}(\vec{k}){\bf \Psi}_1))$ thus 
implying that $({\hat a}_D(\vec{k}))^{\dagger} ={\hat a}_D^{\dagger}(\vec{k})$ on ${\cal H}_{phys}$.

\subsubsection{Equivalence with the standard Fock representation.}

Recall that in the standard Fock quantization, the field operator
${\hat \phi}(X)$ is given by  the expression
\be
{\hat \phi}(X) = 
\frac{1}{(2\pi)^{\frac{n}{2}}}\int \frac{d^nk}{\sqrt{2k}}
 ({\hat {\bf a}}({\vec k}) e^{-ikT +i {\vec k}\cdot{\vec{X}}} + {\hat {\bf a}}^{\dagger}(\vec{k}) e^{ikT -i {\vec k}\cdot{\vec{X}}}),
\label{phixt}
\ee
where we have used the notation $X^0 = T$ and $\{X^A, A=1,..,n\} ={\vec X}$. The operators 
${\hat {\bf a}}({\vec k}), {\hat {\bf a}}^{\dagger}(\vec{k})$ are the annihilation and creation operators on the 
standard Fock space ${\cal F}$. The standard Fock vacuum is denoted by $\arrowvert 0\rangle$ so 
that ${\hat {\bf a}}({\vec k})\arrowvert 0\rangle = 0 \forall {\vec k}$ and ${\cal F}$ is generated by the (repeated) 
action of the creation operators ${\hat {\bf a}}^{\dagger}(\vec{k})$ on $\arrowvert 0\rangle$.
We denote the pair ${\hat {\bf a}}({\vec k}), {\hat {\bf a}}^{\dagger}(\vec{k})$ by $\vec{{\hat {\bf a}}}$.
\footnote {In the notation employed in section 4B, the operators  ${\vec{{\hat {\bf a}}}}$ 
would be denoted by ${\vec{\hat a}}_H(F_0)$ since they are the Heisenberg picture representatives of the classical
variables ${\vec a}(X)$ on the initial slice $X^A(x) = F_0^A(x)$. This notation (and a suitable notation for the 
individual annihilation and creation operators) is cumbersome and this why we have used the simpler notation with 
the bold face.\label{f5}}

We shall now show that this quantization is equivalent to the Dirac quantization of PFT constructed in this work.
We proceed in two steps. First we show that $\arrowvert F_0\rangle \otimes {\cal H}_{F_0}$ is naturally
isomorphic to ${\cal H}_{phys}$. Next we show that $\arrowvert F_0\rangle \otimes {\cal H}_{F_0}$ is also 
naturally isomorphic to ${\cal F}$ and that the consequent isomorphism between 
${\cal F}$ and ${\cal H}_{phys}$ renders the two representations equivalent.

Equations (\ref{2}) and (\ref{2.1}) imply that $\eta$ restricted to $\arrowvert F_0\rangle \otimes {\cal H}_{F_0}$
is bijective. Specifically, 
(\ref{2}) and (\ref{2.1}) imply that every 
$\arrowvert \psi\rangle \in \arrowvert F_0\rangle \otimes {\cal H}_{F_0}$ is uniquely mapped to 
${\bf \Psi}\in {\cal H}_{phys}$ via $\eta \arrowvert \psi\rangle ={\bf \Psi}$.
Further, given $\arrowvert \psi\rangle, \arrowvert \phi\rangle \in \arrowvert F_0\rangle \otimes {\cal H}_{F_0}$,
we have that 
\be
(\eta \arrowvert \psi\rangle, \eta \arrowvert \phi\rangle ) 
= \langle\phi \arrowvert \psi \rangle ,
\ee
\be
{\hat a}_D({\vec k}) \eta \arrowvert \phi\rangle
=\eta {\hat b}_{F_0}({\vec k}) \arrowvert \phi\rangle,
\ee
\be
{\hat a}^{\dagger}_D({\vec k}) \eta \arrowvert \phi\rangle
=\eta {\hat b}^{\dagger}_{F_0}({\vec k}) \arrowvert \phi\rangle .
\ee
These equations imply that $\eta$  defines a 1-1, antilinear map from $\arrowvert F_0\rangle \otimes {\cal H}_{F_0}$
to ${\cal H}_{phys}$ which induces an antilinear $\star$- homeomorphism from the $\star$- algebra of operators 
generated by 
${\hat b}_{F_0}({\vec k}),{\hat b}^{\dagger}_{F_0}({\vec k})$ to that  generated by 
${\hat a}_D({\vec k}), {\hat a}^{\dagger}_D({\vec k})$.
\footnote{The operators which generate the algebra are related by adjointness relations. These relations induce
adjointness relations among elements of the algebra in the obvious way. Such an algebra with adjointness relations
is called a $\star$- algebra.  A $\star$- homomorphism between two $\star$- algebras preserves both the algebraic and
the adjointness relations between elements of the algebras. For a more precise definition of 
these structures, see, for example, Reference \cite{alm2t}.\label{f6}}


Next, note that the representation of the $\star$- algebra of operators generated by 
${\hat b}_{F_0}({\vec k}),{\hat b}^{\dagger}_{F_0}({\vec k})$ on $\arrowvert F_0\rangle \otimes  {\cal H}_{F_0} $
is (trivially,)  unitarily related to that generated by  
${\bf {\hat a}}({\vec k}),{\bf {\hat a}}^{\dagger}$ on ${\cal F}$. Explicitly, we define this 
relation via the map $U:{\cal F}\rightarrow\arrowvert F_0\rangle \otimes {\cal H}_{F_0}$ where
\be 
U\arrowvert 0 \rangle = \arrowvert F_0\rangle \otimes\arrowvert 0,F_0 \rangle ,
\ee
\be
U\prod_{i}^m({\hat {\bf a}}^{\dagger}(\vec{k_i}))^{n_i}\arrowvert 0 \rangle = 
\prod_{i}^m({\hat b}_{F_0}^{\dagger}(\vec{k_i}))^{n_i}\arrowvert F_0\rangle \otimes\arrowvert 0,F_0 \rangle .
\ee
Clearly, we have that
\footnote{ Equation (\ref{a=cb}) implies that ${\vec {\hat b}}(F_0) = {\vec {\hat a}}_S$. The discussion 
in sections 4B, 4C suggests that ${\vec {\hat a}}_S$ be associated with the ``time zero'' Heisenberg 
operators ${\vec {\hat a}}_H(F_0)$. The discussion  after (\ref{phixt}) with regard to notation implies
this association is consistent with equation (\ref{I}) below.\label{f7}}
\be
U \vec{\hat {{\bf a}}} U^{\dagger} = {\vec {\hat b}}(F_0) .
\label{I}
\ee

It follows that the map ${\cal U}:= \eta\circ U$ is a 1-1, antilinear map from ${\cal F}$ to ${\cal H}_{phys}$
which induces an anti- linear $\star$- homomorphism from the $\star$- algebra of operators generated by 
${\bf {\hat a}}({\vec k}),{\bf {\hat a}}^{\dagger}({\vec k})$ to that generated by 
${\hat a}_D({\vec k}), {\hat a}^{\dagger}_D({\vec k})$. In language shorn of mathematical jargon,
all we are saying is that  the Dirac quantization constructed here is
unitarily equivalent to the conjugate  representation defined by the Fock representation, where, by conjugate representation 
we mean the representation defined on `bra' states from that defined by `ket' states.
This completes our proof  of the physical equivalence of the standard Fock quantization and 
the Dirac quantization described in this paper.

\subsection{The functional Schrodinger equation in 1+1 dimensions.}

In this section, our aim is to reveal the structures in the Dirac quantization of PFT which allow for the 
possibility of defining functional Schrodinger equations in 1+1 dimensions of the type considered in
References \cite{karel1+1q,tv1}. We shall assume familiarity with the contents and results of 
References \cite{karel1+1q,tv1} and limit our exposition to a rephrasing of those results in terms of the structures defined
here.

We start with a few remarks which lead to a caveat regarding the applicability of the Torre- Varadarajan analysis to
the case of non- compact Cauchy slices considered here.
Torre and Varadarajan  (TV) \cite{tv1} showed that in 1+1 dimensions, free scalar field evolution from 
an initial flat slice to an arbitrary slice of flat spacetime is unitarily implemented in the standard
Fock space quantization. They defined  Schrodinger picture states by the action of 
the inverse (embedding- dependent) unitary transformation on (embedding- independent) states 
in the standard Fock space. By functionally differentiating the 
Schrodinger picture states with respect to the embedding, TV showed that these states satisfy a functional 
Schrodinger equation which corresponds to a rigorous definition of equation (\ref{fsch}) in quantum theory.
This rigorously defined Schrodinger equation was anticipated in full detail by Kucha{\v r} in \cite{karel1+1q}.
The TV results in 1+1 dimensions \cite{tv1} were derived for compact Cauchy slices diffeomorphic to the circle.
In this paper, for the case of 1+1 dimensions, 
 we have considered non- compact Cauchy slices diffeomorphic to the real line. Hence, strictly speaking,
the TV results do not apply here. However, TV have persuasively argued in section V of Reference \cite{tv1}
that their results should also apply to the case of the spatial topology being that of the line. 
\footnote{Conversely, we believe that our general considerations here should apply with suitable modifications
to the case of compact Cauchy slices.\label{f8}} 
Therefore we shall simply assume that the relevant TV structures exist in the case of the non- compact 
Cauchy slices considered here. 
It is beyond the scope of this paper to provide a detailed generalisation of  the TV analysis
to the non- compact case. We now proceed to a rephrasing of
the (assumed) TV results in terms of our constructions in this work.

Note that if evolution between $F_0^A(x)$ and any $F^A(x)\in {\cal E}$ is unitary on ${\cal F}$, it follows that 
all the Hilbert spaces ${\cal H}_F,\; F^A(x)\in {\cal E}$ are identical. We denote this single 
Schrodinger picture Hilbert space by ${\cal H}$. Of course ${\cal H}= {\cal F}$; we use a different notation 
only to remind the reader that we are dealing with a Schrodinger picture representation.

Next, note that if ${\hat P}_A(x)$ was a well defined operator on ${\cal H}_{kin}$, 
we could directly evaluate its action on ${\bf \Psi} \in {\cal H}_{phys}$ via the dual action 
(\ref{dualaction}). However, as mentioned earlier, besides the problems described in section 5,
the operator ${\hat H}_G$ does not have the requisite continuity in $G^A(x)$ to define
${\hat P}_A(x)$. Specifically, the limit
\be
\lim_{\epsilon\rightarrow 0} \frac{{\hat H}_{\epsilon G}-1}{i \epsilon}\arrowvert \psi \rangle  
:= \int dxG^A(x) {\hat P}_A(x)\arrowvert \psi \rangle
\ee
for $\arrowvert \psi \rangle \in {\cal D}_X$ does not exist on the Hilbert space ${\cal H}_X$ (see section 3C for 
a discussion of ${\cal D}_X, {\cal H}_X$). Although this limit does not exist on ${\cal H}_X$, we show below that such a 
limit can be defined on a suitable subspace of the algebraic dual space ${\cal D}^*$ (see ${\bf (iii)}$ of section 6A
for the definition of ${\cal D}^*$).

Let ${\bf \Psi}=\sum_{{\bar F}\in {\cal E}}
\langle\psi , {\bar F}\arrowvert \otimes\langle  {\bar F}\arrowvert \;\in {\cal H}_{phys}$. We define the `partial' action of
${\bf \Psi}$ on $\arrowvert F\rangle, \; F^A(x)\in {\cal E}$ as follows:
\bea 
{\bf \Psi}(\arrowvert F\rangle )
&:= &(\sum_{{\bar F}\in {\cal E}}\arrowvert {\bar F}\rangle \otimes\arrowvert \psi , {\bar F}\rangle ,
                                                 \arrowvert F\rangle)
\\
&=&  \sum_{{\bar F}\in {\cal E}} \delta_{{\bar F},F}\langle\psi , {\bar F}\arrowvert 
\\
&=&
\langle\psi ,  F\arrowvert .
\eea
Next, consider the action on ${\bf \Psi}$,  of the operator 
$\frac{{\hat H}_{\epsilon G}-1}{i \epsilon}$, in the context of this partial action.
We have that 
\bea 
(\frac{{\hat H}_{\epsilon G}-1}{i \epsilon} {\bf \Psi})(\arrowvert F\rangle )
&=& {\bf \Psi}(\frac{{\hat H}^{\dagger}_{\epsilon G}-1}{-i \epsilon}\arrowvert F\rangle )
\\
&=&
{\bf \Psi}(\frac{\arrowvert F+\epsilon G\rangle -\arrowvert F\rangle}{-i\epsilon}).
\\
&=&
\frac{\langle\psi ,  F+\epsilon G\arrowvert - \langle\psi ,  F\arrowvert}{-i\epsilon}
\label{A1}
\eea
Note that in general $F^A(x) +\epsilon G^A(x) \notin {\cal E}$ for arbitrary $F^A(x), G^A(x),\epsilon$.
However, for a given $G^A(x)$ of compact support and $F^A(x) \in {\cal E}$, it is straightforward to see that 
there exists $\epsilon_0 >0$ such that $\forall \epsilon$ satisfying the condition $0<\epsilon <\epsilon_0$, 
we have that $F^A(x) +\epsilon G^A(x) \in {\cal E}$. 
\footnote{We expect that a  similar result should be true for $G^A(x), F^A(x)$ satisfying suitable `asymptotically flat' conditions
which allow for more general $G^A(x)$.\label{f9}}
Thus, equation (\ref{A1}) is well defined for sufficiently small $\epsilon$.
Note that such an equation would not be meaningful if ${\cal H}_{F+\epsilon G}$ and ${\cal H}_F$ were not identical.
This is expected to happen in higher dimensions for generic choices of $F^A(x), G^A(x)$ \cite{tv2}. Thus equation (\ref{A1}) is expected to
only 
make sense in 1+1 dimensions or in the context of special choices of $F^A(x), G^A(x)$ in higher dimensions.

The $\epsilon \rightarrow 0$  limit of equation (\ref{A1}) exists
on the Hilbert space ${\cal H}$ provided  
$\arrowvert \psi, F+\epsilon G\rangle$ is sufficiently well behaved in $\epsilon$. 
If this happens we may write the $\epsilon \rightarrow 0$ limit of (\ref{A1}) as 
\be
\lim_{\epsilon \rightarrow 0}(\frac{{\hat H}_{\epsilon G}-1}{i \epsilon} {\bf \Psi})(\arrowvert F\rangle )
=i  \int dx G^A(x)\frac{\delta \;\;\;\;}{\delta F^A(x)}
                       \langle\psi ,  F\arrowvert . 
\label{D}
\ee
As we argue below such well- behavedeness should follow from a  specification of the Schrodinger picture 
vacuum state along the lines of Reference \cite{tv1}.
 Recall that we defined $\arrowvert 0,F\rangle$ to be the state annihilated by the 
operators ${\hat b}_F({\vec k}), \forall {\vec k}$.  This definition only specifies 
$\arrowvert 0,F\rangle$ upto an $F^A(x)$ dependent phase. For an arbitrary choice of this phase, the 
$\epsilon \rightarrow 0$  limit of equation (\ref{A1}) may not exist.

In \cite{tv1},  the  embedding dependent- Schrodinger picture- vacuum is constructed by the action of the exponential of an  operator,
quadratic in the creation operators,
on  the standard Fock vacuum, 
multiplied by 
 an explicitly  defined embedding dependent normalization factor. The coefficients of
the creation operators in the exponent are constructed out of the Bogoliubov coefficients 
which define the classical canonical transformation corresponding to (inverse) evolution from the embedding back to the 
initial flat slice.
These structures are in correspondence with the following structures in this work. Clearly the creation operators correspond to
${\bf a}^{\dagger}({\vec k})$ and the standard Fock vacuum to $\arrowvert 0 \rangle$.
Footnote \ref{f6}, together with the fact that ${\cal H}={\cal F}$, implies the identifications
${\hat b}^{\dagger}_{F_0}(\vec{k})= {\hat a}_S^{\dagger} (\vec{k}) = {\bf a}^{\dagger}(\vec{k})$ and
$\arrowvert 0 , F_0\rangle = \arrowvert 0 \rangle$ (in fact, we fix the phase ambiguity in the definition of 
$\arrowvert 0 , F_0\rangle$ by this identification). The Bogoliubov transformation is $C(F,F_0)$ and the 
Bogoliubov coefficients are 
$\alpha_{F,F_0}(\vec{k}, \vec{l}),\beta_{F,F_0}(\vec{k}, \vec{l})$ where ${\vec k},{\vec l}$ are in correspondence with 
the set of real numbers since they are spatial vectors in 1 dimension
(see  equation (\ref{cf2f1}) for the definition of $\alpha , \beta $).

We expect that, as in the spatially compact case,  
the Schrodinger picture vacuum 
can be explicitly constructed  
along the lines of \cite{tv1} and that, as in \cite{tv1}, 
it is functionally differentiable with respect to the embedding. Thus, we expect that equation (\ref{D}) is well defined
for $\arrowvert \psi , F\rangle =\arrowvert 0 , F\rangle$.  The Schrodinger picture $N$- particle states of section 5C can then
be obtained by the repeated action of the creation operators ${\hat b}^{\dagger}_F(\vec{k})$ on $\arrowvert 0 , F\rangle$.
From \cite{tv1} we expect that these states should also be functionally differentiable with respect to $F^A(x)$. It follows
that any $\arrowvert \psi , F\rangle$ which is a suitably well behaved linear combination of these basis states is also 
functionally differentiable. The right hand side of  equation (\ref{D}) can be explicitly computed for 
such states in terms of the functional derivatives of the basis states. We expect that the latter can be computed
along the lines of \cite{tv1} and that here,
in close analogy to
the result of such a computation in Reference \cite{tv1}, 
$\langle\psi ,  F\arrowvert$ satisfies a rigorously defined functional Schrodinger equation with a precisely defined, 
non- trivial operator ordering prescription for ${\hat h_A}(x)$ and an additional $c$- number correction.
\footnote{ The expected  functional Schrodinger equation may be identified with a definition of the constraint operator ${\hat C}_A(x)$ on
${\cal H}$. As in  Reference \cite{karel1+1q,tv1}, the  $c$- number correction is expected to compensate for a Virasoro
type anomaly in the constraint algebra which arises if ${\hat C}_A(x)$ is defined without this additional correction; it is 
only with this correction that the constraint algebra is expected to close \cite{karel1+1q}.\label{f10}}

This completes our discussion of the (expected) derivation of the 1+1 dimensional functional Schrodinger equation 
from the Dirac quantization constructed in this paper.

\section{Discussion.}
 In particle quantum mechanics, the Heisenberg picture is identified with the Schrodinger picture at some initial instant of time.
In the case of $n+1$ dimensional PFT, Torre and Varadarajan (TV) choose this identification to be at an initial flat slice
defined by $X^A(x) = (0, x^1,..,x^n)$. Their results imply that embedding dependent Schrodinger picture states 
cannot be constructed as the unitary images of (embedding independent)  states in the standard (Heisenberg picture)
Fock space, ${\cal F}$, for generic choices of embedding and for $n>1$. This rules out the possibility of 
defining (unitary) functional evolution of states on ${\cal F}$.
\footnote{Since embeddings are specified by functions- worth of data, 
we refer to the evolution from the initial slice to an arbitrary final one as `functional' evolution.\label{f11}}

While such evolution of states in the single Hilbert space ${\cal F}$ is ruled out, TV showed that functional evolution
of states {\em can} be defined, provided the notion of a quantum state is enlarged to that used in algebraic quantum field theory
\cite{tv2}. There, algebraic states are identified with positive linear functionals (PLFs) on the Weyl algebra (see, for example,
\cite{bogbovcondtns}). Conversely, every  PLF on the Weyl algebra defines, through the Gelfand- Naimark- Segal (GNS) construction
\cite{bogbovcondtns}, a Hilbert space representation of the Weyl algebra. This representation is such that there is a state in the 
GNS Hilbert space for which the expectation value of any Weyl algebra element is the same as the evaluation of the PLF 
on this element. Consider the Weyl algebra for a free scalar field on flat spacetime. Inverse evolution from any embedding to the
initial flat embedding is a canonical transformation and, hence, defines an automorphism of the Weyl algebra \cite{tv2}. The 
induced action of this automorphism on any PLF defines a new PLF i.e. a new algebraic state. Any state in 
${\cal F}$ defines, via its expectation values in the standard Fock representation, 
a PLF on the Weyl algebra. The action of the automorphism corresponding to
inverse evolution from an arbitrary embedding to the initial flat one on this PLF yields an embedding dependent PLF i.e.
an embedding dependent algebraic state. 
It is in this sense that functional evolution of algebraic states is well defined.

In the context of this algebraic quantum field theory viewpoint, the TV results indicate that the 
GNS Hilbert space representation associated with a generic embedding is  inequivalent 
to the Fock space representation. Indeed, the embedding dependent Hilbert spaces ${\cal H}_F$ defined in section 4D
are precisely the embedding dependent GNS Hilbert spaces.
\footnote{Denote the GNS Hilbert space associated with $F^A(x)$ by ${\cal H}^{(GNS)}_F$. Then 
${\cal H}^{(GNS)}_F$ is naturally dentified with ${\cal H}_F$ in such a way that  
${\hat {\bf a}}(\vec{k}),{\hat {\bf a}}^{\dagger}(\vec{k})$ are mapped to ${\hat a}_S({\vec k}),{\hat a}^{\dagger}_S({\vec k})$.
See Footnote \ref{f7} with regard to the naturalness of this identification.\label{f12}}
Thus, our work can be considered to be an implementation, using LQG techniques, of algebraic state evolution
in the context of the single , non- seperable Hilbert space ${\cal H}_{kin}$  (\ref{defhkin}).  Moreover, even though
generic scalar field evolution is not unitary on ${\cal F}$, the evolution generated by the constraints, as defined by
${\hat U}_d$, {\em is}
unitary on the ``much larger'' Hilbert space ${\cal H}_{kin}$ (see section 5B). Clearly, the LQG type of representation
of the embedding variables plays a key role in this unitarity.

Despite our demonstration that the TV results are not an obstruction to a Dirac quantization of PFT, 
these results {\em can} be restated in the context of Dirac quantization as follows. Rather than interpret them
in the context of a (putative) Schrodinger picture, the TV results can be interpreted within the context of 
Heisenberg picture based quantizations. In this context, they imply the existence of inequivalent 
quantizations of the free scalar field. These arise from  inequivalent complex structures i.e. inequivalent
choices of basic annihilation and creation operators. Specifically, given an embedding $F^A_1(x)$, the operators
${\vec{{\hat {\bf a}}}}_1 =C(F_0,F_1) \vec{{\hat {\bf a}}}$ (see equation (\ref{phixt}) for the definition of $\vec{{\hat {\bf a}}}$)
are the annhilation and creation operators for the Hilbert space ${\cal H}_{F_1}$.
Any embedding dependent Heisenberg picture  operator can be constructed from 
the operators corresponding to $\vec{{\hat {\bf a}}}$, and then re- expressed in terms of the operators 
${\vec{{\hat {\bf a}}}}_1$.
If the Bogoliubov 
transformation $C(F_0,F_1)$ is not unitarily implementable on ${\cal F}$ then the Heisenberg picture representation on ${\cal H}_{F_1}$ is
inequivalent to the standard Heisenberg picture Fock representation on ${\cal F}$.
This unitary inequivalence can be traced to the fact that the operator ${\hat N}_0$ 
which measures the number of excitations associated with
$\vec{{\hat {\bf a}}}$ (i.e. 
${\hat N}_0=\int  d^nk {\hat {\bf a}}^{\dagger}(\vec{k}){\hat {\bf a}}(\vec{k})$)
has a well defined  vacuum expectation value in ${\cal F}$ but not in 
${\cal H}_{F_1}$ (where the `vacuum' state in  ${\cal H}_{F_1}$ is defined to be $\arrowvert 0,F_1 \rangle$) \cite{bogbovcondtns}.
 Conversely the number operator ${\hat N}_1$ for the
excitations associated with the operators ${\vec{{\hat {\bf a}}}}_1$ has a  well defined vacuum expectation value in 
${\cal H}_{F_1}$ but not in ${\cal F}$.

A Dirac quantization of PFT, equivalent to this quantization, can be constructed by replacing the 
operators $\vec{\hat b}(F)$ by (in obvious notation) the operators ${\vec{\hat h}}(F)= C(F_0, F_1)\vec{\hat b}(F)$,
\footnote{In order not to have too cumbersome a notation, we have ommitted to signify the dependence of ${\vec{\hat h}}(F)$
on the fixed embedding $F_1^A(x)$.\label{f13}} 
and the Dirac observables ${\vec {\hat a}}_D$ by  the Dirac observables  ${\vec {\hat a}}_{(1)D}= C(F_0, F_1) {\hat a}_D$,
in our constructions. Thus, the new kinematic Hilbert space, ${\cal H}^{1}_{kin}$, is given by
${\cal H}^{1}_{kin} = \bigoplus_F \arrowvert F\rangle \otimes{\cal H}^{1}_F$ where 
${\cal H}^{1}_F$ is the Fock space associated with the annhilation and creation operators 
${\hat h}({\vec k}),{\hat h}^{\dagger}({\vec k})$. The interested reader can check that all the steps go through and 
one obtains a quantization equivalent to the Heisenberg picture quantization on ${\cal H}_{F_1}$.
In this quantization the Dirac observable  ${\hat N}_{(0)D}= \int  d^nk {\hat a}_D^{\dagger}(\vec{k}){\hat a}_D(\vec{k})$
does not have a  well defined  vacuum expectation value  (where the vacuum is defined to be the state annihilated by
 ${\hat a}_{(1)D}({\vec k})$) whereas 
the Dirac observable 
${\hat N}_{(1)D}= \int  d^nk {\hat a}_{(1)D}^{\dagger}(\vec{k}){\hat a}_{(1)D}(\vec{k})$  does have a well defined (vanishing)
vacuum expectation value.
Conversely ${\hat N}_{(1)D}$ does not have a well defined vacuum expectation value in the representation constructed in section 6 
(where the vacuum is defined to be the state annihilated by ${\hat a}_{D}({\vec k})$) whereas the vacuum expectation value of
${\hat N}_{(0)D}$ vanishes. This concludes our discussion of the relation of the TV results to the constructions of this work.

We shall discuss the very interesting open issue regarding the existence and structure of 
${\cal G}({\cal E})$ towards the end of this section.
Now, we turn to a summary of technical details which need to be worked out in the Dirac quantization of PFT. 
As mentioned in 
section 2B, we have not specified asymptotic boundary conditions on the embedding variables.
As mentioned in Footnote \ref{f4}, we have been cavalier about the distinctions between genuine operators and
operator valued distributions, as well as those between bounded and unbounded operators.
We feel that these details can easily be supplied in a more careful treatment and that our results
will be unaffected. In section 6D, we assumed that our results would also apply, in the 1+1 dimensional case,
to spatially compact slices of topology $S^1$. In the spatially compact case, there are two new ingredients.
The first is the absence of global inertial coordinates. This should easily be handled, as in \cite{karel1+1c},
by describing the embedding variables as suitable maps from $S^1$ to the spacetime. The second new ingredient is 
the appearance of `zero modes' of the scalar field; these are quantum {\em mechanical} (as opposed to
field theoretic) degrees of freedom and we expect that they can be accomodated without changing our basic results, but
this needs to be worked out. 
Note also that, in the case studied in this work (i.e. the case of the spatial topology being that of $R^n$), 
we have not explicitly worked out the Bogoliubov coefficients. As noted in \cite{tv1}, there may be infrared 
problems when $n=1$- we refer the reader to the comments in \cite{tv1} as they apply here as well.
Note that in this case (i.e. of the spatial topology being that of a line), a Dirac quantization of PFT
different from ours, has been recently constructed by Laddha \cite{alok}. It would be of interest to 
compare his quantization with ours.

Let us assume that our results do go through as envisaged in 1+1 dimensions for the spatially compact case. 
Then Footnote \ref{f10} indicates that 
despite an anomaly free quantization of the constraints, the formalism is still sensitive to the Virasoro anomaly.
Further, as stressed by Kucha{\v r} in \cite{karel1+1q}, although there is no anomaly in the algebra of constraints
due to the compensating `anomaly potential' term, there {\em is} the usual Virasoro anomaly in the 
algebra of Dirac observables which correspond to the 
normal ordered stress energy fluxes.
Since the closed bosonic string can be realised as a PFT with an additional constraint \cite{charliekarelstring},
we believe  that a Dirac quantization, along the lines sketched here, of this system should be possible and should yield
a quantization identical to that of \cite{charliekarelstring}. It would be of interest to attempt to construct such a quantization
and compare it with Thiemann's quantization of the string \cite{thomas}.

We also believe that our constructions here can be suitably ( and trivially) modified so as to apply to the case of axisymmetric
PFT in 2+1 dimensions. Thus, we expect that despite the results of \cite{chome}, it is possible to construct a Dirac quantization 
of the system which is equivalent to the standard Fock quantization. 
The results of \cite{tv2,chome} obtained in the context of higher dimensional PFTs and cylindrical waves seemed to raise apprehensions
about the physical viablity of any Dirac quantization based approach to quantum gravity. It is ironical 
that in the light of our results  here,  the Dirac quantization of these  systems
(certainly PFTs and most likely, cylindrical waves, in their reformulation as axisymmetric PFT) are  beautiful examples of the 
power of the techniques developed in one  such approach, namely that of LQG.

As mentioned earlier in this section, we shall now turn to a discussion of the most interesting open issue in this work, namely that
of a precise characterisation of the space of finite canonical transformations generated by the smeared constraints
$C(\xi)$ of equation (\ref{defcxi}). Recall that the problem is that while the Poisson bracket algebra of the smeared constraints
is isomorphic to the Lie bracket of vector fields on the spacetime, the finite canonical transformations generated by 
$C(\xi)$ are not in correspondence with the group of all spacetime diffeomorphisms because the latter do not leave the 
space of spacelike embeddings, ${\cal E}$, invariant. It is therefore of interest to know whether there exists an 
infinite dimensional space of diffeomorphisms which keep  ${\cal E}$ invariant. If this (putative) space exists as the group
${\cal G}({\cal E})$ of section 5A, and if ${\cal E}$ can be generated by the action of ${\cal G}({\cal E})$ on any fixed 
element of ${\cal E}$, the technique of Group Averaging yields the Dirac quantization constructed in this work. 
If ${\cal G}({\cal E})$ does not have this property, we expect the existence of superselection sectors in the space
of physical states. Clearly, the detailed structure of ${\cal G}({\cal E})$ (if it exists) determines the structure of the 
physical Hilbert space.

Since the work of \cite{karelisham} seems to indicate that ${\cal G}({\cal E})$ may not exist as an infinite dimensional group, 
we have replaced the 
set of finite canonical transformations generated by the smeared constraints by the (intuitively) much larger set of
transformations labelled by bijections from ${\cal E}$ to itself i.e. by elements of ${\cal S}({\cal E})$.
As discussed in section 5B,  such bijections do not have any continuity properties reflective of the differential structure of the
spacetime manifold $M$. Nevertheless, this `enlargement' of the set of gauge transformations in quantum theory yields the
physically correct quantization on a {\em seperable} Hilbert space. It would be useful to understand exactly why this 
happens. This enlargement of the notion of gauge is reminiscent of the proposed enlargement of the spatial diffeomorphism
gauge in LQG by Zapata \cite{zapata}. Zapata showed that his proposal lead to a seperable Hilbert space prior to the 
imposition of the Hamiltonian constraint. It is not clear what repercussions Zapata's proposed `enlargement' of spatial
diffeomorphism gauge has on the classical limit of LQG. By virtue of the close analogy between structures in the Dirac quantization
of PFT and in LQG, we hope that this work may be of some use in clarifying the above issue as well as other issues (such 
as those of interpretation) in LQG.
Note that one key difference between the PFT case and LQG is that in the former, the smeared constraints form
a Lie algebra. It may be of interest to see if progress can be made in an LQG type of quantization of more complicated systems
in which the constraint algebra is a Lie algebra, such as that of gravity with appropriate matter \cite{karelcommute}.

\vspace{2mm}

\noindent {\bf Acknowledgements}: We thank Abhay Ashtekar for very useful conversations about this work.
We thank Charles Torre for very enjoyable interactions over the years, for his constant encouragement
and for very useful conversations about this work.



\end{document}